\documentclass[aps,prb,showpacs,superscriptaddress,floatfix,twocolumn,10pt]{revtex4-1}
\usepackage{graphicx, color}
\usepackage{amsmath, amsthm, amssymb}
\usepackage{bm}
\usepackage{hyperref}
\usepackage{overpic}
\newcommand{\mb}[1]{\mathbf{#1}}
\newcommand{\pd}{\partial}

\newcommand{\up}{\uparrow}
\newcommand{\down}{\downarrow}
\newcommand{\mbb}[1]{\mathbb{#1}}
\newcommand{\mc}[1]{\mathcal{#1}}

\newcommand{\hc}{{\rm H.c.}}

\def\e{\epsilon}
\def\la{\langle}
\def\ra{\rangle}
\def\sgn{\textrm{sgn}}

\begin{document}
\title{Persistent currents in Dirac fermion rings}

\author{Doru Sticlet}
\email{sticlet@pks.mpg.de}
\affiliation{LOMA (UMR-5798), CNRS and University Bordeaux 1, F-33045 Talence, France}
\affiliation{Max-Planck-Institut f\"ur Physik komplexer Systeme, N\"othnitzer Str. 38, 01187 Dresden, Germany}

\author{Bal\'azs D\'ora}
\affiliation{Department of Physics and BME-MTA Exotic Quantum Phases Research Group, Budapest University of Technology and  Economics, Budafoki \'ut 8, 1111 Budapest, Hungary}

\author{J\'er\^ome Cayssol}
\email{jerome.cayssol@u-bordeaux1.fr}
\affiliation{LOMA (UMR-5798), CNRS and University Bordeaux 1, F-33045 Talence, France}
\affiliation{Max-Planck-Institut f\"ur Physik komplexer Systeme, N\"othnitzer Str. 38, 01187 Dresden, Germany}

%\date{\today}

\begin{abstract}
The persistent current in strictly one-dimensional Dirac systems is investigated within two different models, defined in the continuum and on a lattice, respectively. The object of the study is the effect of a single magnetic or nonmagnetic impurity in the two systems.
In the continuum Dirac model, an analytical expression for the persistent current flowing along a ring with a single delta-like magnetic impurity is obtained after regularization of the unbounded negative energy states.
The predicted decay of the persistent current agrees with the lattice simulations. The results are generalized to finite temperatures. To realize a single Dirac massless fermion, the lattice model breaks the time-reversal symmetry, and in contrast with the continuum model, a pointlike nonmagnetic impurity can lead to a decay in the persistent current.

 \end{abstract}
\maketitle
\section{Introduction}

An isolated normal-metal ring threaded by a magnetic flux $\Phi$ carries a nondissipative current $I(\Phi)$ at very low temperature.\cite{Imry2008,AkkM2011}
This is called the persistent current (PC) and it is a manifestation of quantum mechanics, reminiscent of orbital magnetism in atomic physics. At zero-temperature, $I(\Phi)$ is defined as the change in the ground-state energy of the electronic fluid with respect to the magnetic flux. According to gauge invariance, $I(\Phi)$ is a periodic function of the magnetic flux,\cite{Byers1961} the period being the magnetic flux quantum $\Phi_0 = h/e$, where $h$ is the Planck constant and $-e$, the electron charge.
The PC survives at finite temperature, and in the presence of static disorder, as long as the electronic wave functions are coherent over the whole ring.\cite{Buettiker1983} The current-flux relation $I(\Phi)$ is sensitive to many physical ingredients including band structure, static disorder, electronic interactions,\citep{Ambegaokar1990,Berkovits:1993,Muller:1994, Bary-Soroker2008} ring geometry, and measurement back-action.
 
The simplest case, namely strictly one-dimensional and noninteracting electrons with quadratic dispersion, has been studied thoroughly by Cheung {\it et al.},\cite{Cheung1988,Cheung1989} both for ballistic and disordered rings. In the clean limit, the zero temperature current-flux relation $I(\Phi)$ is a piecewise linear function of $\Phi$ with discontinuities. The maximal current amplitude is given by $I_0 =e v_F/L$, where $v_F$ is the Fermi velocity and $L$ the ring circumference. In the presence of a single impurity, the discontinuities are rounded, and the maximal current is reduced. Finally, in the presence of multiple scatterers, the PC is further suppressed by the Anderson localization, but remains finite due to the presence of electronic states whose localization length exceeds the ring size.\cite{Cheung1988}  This work has been extended to more realistic situations including additional ingredients such as multichannel effects and/or interaction effects.\cite{Viefers2004}  After decades of controversy between those theoretical predictions and the pioneering experiments,\cite{Levy1990, Chandrasekhar1991, Mailly1993} recent experiments on a single ring and on arrays of many rings agree well with a model based on the noninteracting electrons.\cite{BJ2009,Bluhm2009,Souche2013}

Recent years have seen a surge in studies on materials hosting Dirac fermions, including graphene\cite{CastroNeto2009, Goerbig2011} and topological insulators.\cite{Hasan2010, Qi2011} In graphene, the particular honeycomb lattice implies a linear dispersion of electrons near two isolated Dirac points. Such electrons are described by a two-component spinor wave function corresponding to the sublattice isopin. The two-dimensional (2D) surface states of topological insulators are characterized by a single (or an odd number of) Dirac cone(s) and the electron momentum is locked to the real spin (instead of the lattice isospin, in graphene). Similarly, the one-dimensional (1D) edge state of the quantum spin Hall state is characterized by a single Fermi surface (consisting of only two Fermi points instead of four Fermi points).  

Being a measure of the flux sensitivity of the wave function, the PC is a natural observable to investigate eventual signatures of Dirac physics. 
However, most theoretical (and all experimental) work on PC has been so far focused on nonrelativistic electrons in standard metals, which are often described by a parabolic dispersion relation, and by scalar wave functions.\cite{Cheung1988,Cheung1989}  Notable exceptions are the theoretical investigations of PC in graphene \cite{Recher2007, Zarenia2010} and topological insulator rings\cite{Michetti2011} in the ballistic limit and with realistic finite-width geometry. In rings patterned in HgTe/CdTe 2D topological insulators, the edge states attached to the inner and outer circumferences of the ring overlap, and therefore they are gapped even for relatively large rings. For smaller rings, those hybridized edge states are even pushed in the 2D bulk gap of the HgTe/CdTe well.\cite{Michetti2011} In order to avoid these effects, and to study a single gapless Dirac fermion, one can use the disk geometry.\cite{Michetti2011} This is an interesting possibility offered by topological insulators: one can have Aharonov-Bohm (AB) like effects in singly connected samples because the insulating bulk defines a region where the edge carriers are excluded (as the hole region for a metallic ring).
In the absence of disorder, the shape of the current-flux relation, $I(\Phi)$, turns out to be the same for nonrelativistic scalar electrons \cite{Cheung1988} and for massless Dirac electrons.\cite{Manton1985, Shifman1991, Recher2007, Michetti2011} Besides, the PC flowing along ballistic bismuth rings was calculated assuming that the dispersion is well approximated by the one of a massive Dirac fermion.\cite{Kohno1992} Persistent currents from fermions with linear spectrum were also studied in the context of Luttinger liquid physics. Here one works in a low-energy approximation, where a quadratic spectrum is linearized at the Fermi momentum.\cite{Loss1992a,*Gogolin1994,*Affleck2001,*Eckle2001} In clean systems, the PC obtained in this approximation proved to be identical to the one determined when the entire quadratic spectrum is considered.

Experimentally, the AB oscillations of the conductance in topological insulators have been observed.\cite{Peng2010,Dufouleur2013} Disorder effects turn out to be crucial in determining the nature of the coherent surface state: namely, a single Dirac cone should host a perfectly transmitted mode robust to strong disorder.\cite{Bardarson2010,Zhang2010,Bardarson2013} Such experiments provide hope that the PC could be detected in Dirac materials in a near future.

Motivated by these recent advances in transport, and anticipating future PC experiments on topological insulator rings, the focus of the present article is on the effect of a single impurity in strictly one-dimensional rings. To this aim, we have studied both massless Dirac fermions confined in a continuous loop (Fig.~\ref{fig:AB}), and lattice Dirac fermions hopping on the discrete sites of a ring (Fig.~\ref{fig:Creutz}). For both models, a comparative study of the effect of magnetic or nonmagnetic impurity is undertaken. 

The outline of this paper reads as follows: In Section~\ref{sec:models}, we introduce the Dirac models studied along this paper, and the formula for evaluating the PC-flux dependence, $I(\Phi)$. The first (continuum) model describes one-dimensional, helical, massless Dirac fermions confined in a continuous loop (Fig.~\ref{fig:AB}). The second model is a time-reversal-symmetry-breaking lattice model, implemented on a ring made of discrete sites with spin-dependent hopping terms (Fig.~\ref{fig:Creutz}). The time-reversal-breaking terms allow tuning the 1D band structure from a situation with two flavors of massless Dirac fermions to a situation with a single massless Dirac fermion (Fig.~\ref{fig:dispersion}). 

Section~\ref{sec:cleanring} presents the case of a clean ring, where the continuum and the tight-binding models are compared. First we review the regularization for the continuum Dirac model,\cite{Manton1985, Shifman1991} which leads to a finite total energy and, consequently, to a well-defined persistent current. The lattice model reproduces perfectly the current-flux characteristic, $I(\Phi)$, of the regularized Dirac spectrum in the continuum.

Section~\ref{sec:singleimpurity} treats the case of a single impurity in the ring. In the continuum model,  a nonmagnetic impurity has no effect on the persistent current, because it does not couple the left- and right-movers. Nevertheless, the lattice model lacks time-reversal symmetry (TRS). Hence there is no Kramers degeneracy to protect the crossing in the energy-flux spectrum. Consequently, even nonmagnetic impurities can open up gaps and lead to a suppression of the PC.
In contrast, the case of a magnetic impurity allows a direct comparison between the lattice and continuum model. Because TRS is broken in both models, the same mechanism induces backscattering, thereby producing a decrease of the PC.
The analytical expression obtained in the case of a delta-like impurity is cross-checked with the numerical results revealing a good agreement for any impurity strength, after a renormalization of impurity potential in the continuum model.

Section~\ref{sec:temp} generalizes the results of the previous sections to finite temperatures. The use of the ultraviolet regularization proves particularly useful in determining the persistent current for the case where the temperature and a single magnetic impurity jointly work to decrease the maximum amplitude of the persistent current. The equivalence between the lattice and the continuum models (established at $T=0$) also holds at finite temperature.

In brief, the paper is organized as follows. The Dirac models and the formalism are presented in Sec.~\ref{sec:models}. The main body of the paper is devoted to persistent currents flowing in a Dirac ring with no impurity (Sec.~\ref{sec:cleanring}) and a single impurity (Sec. \ref{sec:singleimpurity}). The generalization to finite temperatures is contained in Sec~\ref{sec:temp}, with additional details enclosed in the Appendix. Conclusions and perspectives are given in Sec~\ref{sec:conclusion}.

\section{Dirac models}
\label{sec:models}
This section introduces the two Dirac models investigated in this paper, and the formalism used to evaluate the persistent current-flux relation, $I(\Phi)$. The first model describes a helical metal defined on a continuous loop (Fig.~\ref{fig:AB}), while the second one is a tight-binding model defined on a ring of discrete sites with lattice spacing $a$ (Fig.~\ref{fig:Creutz}). The two models share the same low-energy spectrum, consisting of a single Dirac branch, over a wide range of parameters. Nevertheless, the dispersions of the higher energy parts of their spectra differ drastically (Fig.~\ref{fig:dispersion}). Indeed the spectrum of the tight-binding model is periodic and bounded, whereas the continuum helical model has an infinite number of negative energy states. Most importantly, in the absence of external magnetic flux ($\Phi=0$), the continuum helical model is time-reversal invariant, whereas the tight-binding model is not, owing to the presence of spin-mixing terms. In the next sections, we will investigate the implications of those differences on the thermodynamical PC, first for a clean ring (Sec. \ref{sec:cleanring}) and then for a ring with an impurity (Sec. \ref{sec:singleimpurity}). 

\subsection{Continuum helical model}
\begin{figure}[t]
\centering
\includegraphics[width=4.5cm]{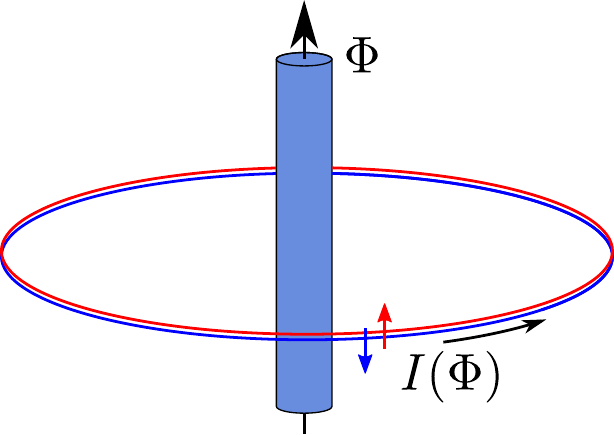}
\caption{(Color online) The helical model Eq.~(\ref{hamiltonianconti}) is implemented on a ring of length $L=2 \pi R$, threaded by a tube of magnetic flux $\Phi$. There are two counter propagating spin $\sigma_3$ eigenstates denoted by $\up$ and $\down$ arrows. 
The arrow in the azimuthal direction indicates the convention for positive (paramagnetic) persistent current $I(\Phi)$.
}
\label{fig:AB}
\end{figure}

Let us consider a strictly one-dimensional metallic ring, lying in the $xy$ plane, with radius $R$, and threaded by a tube of magnetic flux $\Phi$ oriented along the $z$ direction (Fig.~\ref{fig:AB}). It is assumed that the magnetic field is zero for electrons embedded in the ring, and the electronic wave function is modified only through a phase dependence on the electromagnetic vector potential $\mb A = (\Phi / 2 \pi R ) \mb e_\theta$. The Zeeman coupling of the spins with the magnetic field is neglected. 

Let us consider the single-electron Hamiltonian:
\begin{equation}\label{hamiltonianconti}
H=\hbar \omega \bigg(-i\pd_\theta+\frac{\Phi}{\Phi_0}\bigg)\sigma_3,
\end{equation}
where  $\omega=v_F/R$, $v_F$ being the Fermi velocity, and $\hbar=h/2\pi$, the reduced Planck constant. The wave functions are two-component spinors that depend on the azimuthal angle $\theta$ in the $xy$ plane, and $\sigma_3$ is the standard diagonal Pauli matrix. In particular, this model can be seen as an effective model for the helical edge state of the 2D quantum spin Hall insulator occupying a disk with radius $R$ (assuming that the insulator gap is so large that bulk excitations can be neglected). 

Let us look for stationary states under the form $\Psi(\theta,t)=\Psi_n e^{i n \theta-iEt}$, where $\Psi_n$ is a ($\theta$-independent) two-component spinor, and $n$ has to be an integer in order to satisfy periodic boundary conditions $\Psi(\theta,t)=\Psi(\theta+2\pi,t)$.
The eigenstates are proportional to spin eigenvectors, $|z\pm\ra$, of the $\sigma_3$ Pauli matrix: $\sigma_3|z\pm\ra=\pm|z\pm\ra$. 

The two wave functions read
\begin{equation}
\Psi_n^\pm(\theta,t)=|z\pm\ra e^{i n \theta - i E_n^\pm t},
\end{equation}
where $+(-)$ stands for the (anti)clockwise movers. The corresponding energies depend linearly on the magnetic flux
\begin{equation}\label{spectrumconti}
E^\pm_n (\Phi)=\pm\hbar\omega \bigg(n+\frac{\Phi}{\Phi_0}\bigg).
\end{equation}
Therefore the Dirac model in Eq.~(\ref{hamiltonianconti}) has a strong spin-momentum-locking property, clockwise (anticlockwise) states being totally  spin-up (or spin-down) polarized (Fig.~\ref{fig:energyflux}). 

When the flux $\Phi$ is an integer or a half-integer multiple of the magnetic flux quantum, $\Phi_0$, the system is time-reversal invariant, and each single-electron state is twofold degenerated with respect to the spin, ($\sigma_3=\pm 1$). Indeed the action of the time-reversal symmetry $T$ is to reverse both the spin and the momentum operator $-i \partial_\theta$. Therefore the Hamiltonian Eq.~(\ref{hamiltonianconti}) is time-reversal invariant if and only if the flux $\Phi$ is equivalent to flux $-\Phi$, which happens only for integer or half-integer multiples of $\Phi_0$. For arbitrary non-(half-)integer flux $\Phi/\Phi_0$, the external flux breaks TRS, and the spin degeneracy is lifted, as shown by Eq.~(\ref{spectrumconti}) and Fig.~\ref{fig:energyflux}.

Gauge invariance manifests itself in the periodicity of the spectrum with respect to the magnetic flux $\Phi$. Addition of an integer number of flux quanta has no effect on the overall spectrum because the set of energies Eq.~(\ref{spectrumconti}) is unchanged after proper shifting of the integers $n$.

The use of boundary conditions is justified by the fact that the spin quantization axis of the edge state does not depend on the angle $\theta$. This situation is to be contrasted with the case of a surface state wrapping around a topological insulator cylinder, recently investigated in Refs.~\onlinecite{Zhang2010,Bardarson2010,Imura2012,Bardarson2013}. In these works on cylinders, the spin axis winds by $2\pi$, when circulated around the cylinder.

\subsection{Lattice model}
It would be interesting to simulate the single Dirac fermion described by Eq.~(\ref{hamiltonianconti}) by a lattice Hamiltonian for several purposes, including investigations of band-curvature effects at high energy, and temperature effects. Nevertheless a straightforward discretization of a single-flavor Dirac model (as Eq.~(\ref{hamiltonianconti})) is forbidden by the fermion-doubling problem: any time-reversal invariant lattice model has an even number of Dirac points. 

\begin{figure}[t]
\centering
\includegraphics[width=0.60\columnwidth]{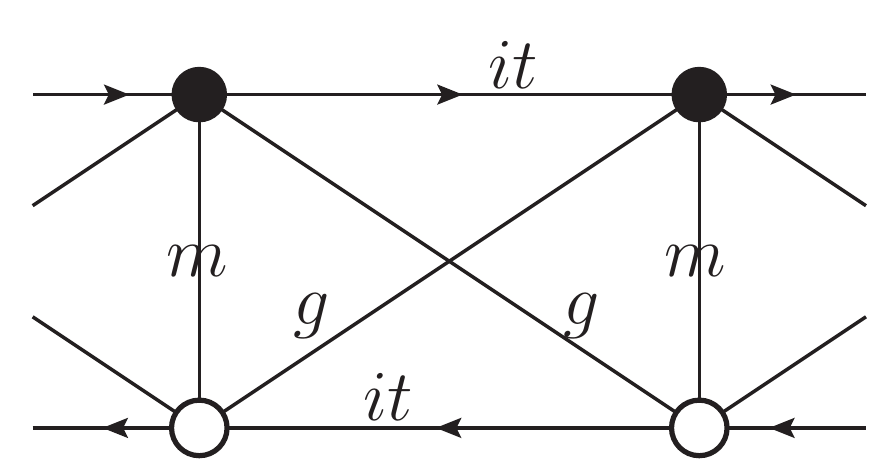}
\includegraphics[width=0.38\columnwidth]{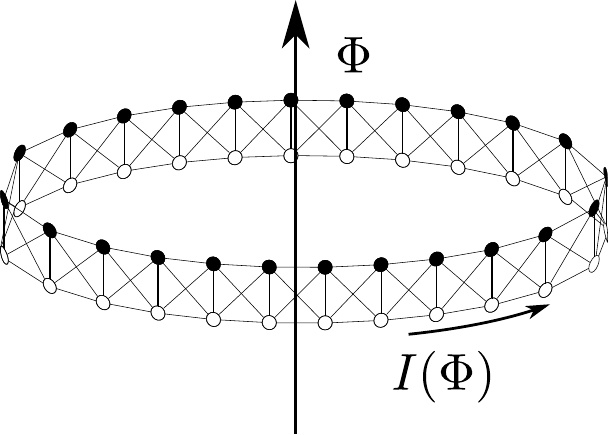}
\caption{Left: Creutz model in zero flux on a linear infinite lattice. The horizontal ladders represent symbolically the two spin projections: the upper (lower) ladder with filled (empty) circles denotes $\sigma_3$ spin-up (spin-down) states. The spin degeneracy is lifted by an on-site coupling $m$. There are two possible hopping amplitudes: the $g$ term that produces spin-flip processes and a spin-conserving hopping $t$. The additional phase $\pi/2$ on the $t$ links is only conventional and has the role to move the Dirac point at $k=0$. Right: The model is shaped into a ring of length $L$ and a flux $\Phi$ is threaded through it. 
}
\label{fig:Creutz}
\end{figure}

This paper focuses on a lattice model that breaks time-reversal symmetry (even at vanishing flux $\Phi$). The single-electron tight-binding Hamiltonian in real space reads
\begin{equation}\label{creutz}
H_{\rm Creutz}=\frac{1}{2}\sum_{j=1}^{L/a}[c^\dag_j(it\sigma_3-g\sigma_1)e^{i \phi } c_{j-1}+mc^\dag_j\sigma_1c_j]+\hc,
\end{equation}
where the spin indices are implied, $c^\dag_j=(c^\dag_{j\uparrow},c^\dag_{j\downarrow})$, and the integer $j\in\{1,\dots,L/a\}$ indexes the different lattice sites (Fig.~\ref{fig:Creutz}). We work in units where the lattice constant is $a=1$. Therefore the wire length $L$ equals the number of sites in the ring. This model has already been studied for other purposes in Refs.~\onlinecite{Creutz1994,*Creutz1999,*Bermudez2009,*Viyuela2012}, and a proposal for its implementation using cold atoms trapped in optical lattices was recently advanced.\cite{Mazza2012} The terms in $\sigma_1$ (proportional to the parameters $m$ and $g$) mix the two spin directions and break time-reversal symmetry even at $\Phi=0$. There is also an orbital effect due to the vector potential, which is reflected in the phase
\begin{equation}
\phi  = \frac{2\pi a}{L}\frac{\Phi}{\Phi_0},
\end{equation}
gained by an electron hopping between nearest-neighboring sites. Note that at $\Phi=0$, there is a phase $\pi/2$ gained by spin-conserving hoppings $t$. This does not break the time-reversal symmetry because a spin-up electron gains the phase $\pi/2$, while the spin-down electron will lose the phase $\pi/2$. In the ring geometry, the sites $j=1$ and $j=N+1$ are identified. In the following simulations, the energies are expressed in units where the hopping strength is dimensionless, $t=1$.

Before considering the model on a finite-size ring, let us discuss its properties on an infinite lattice and in zero magnetic flux. The model is classified in the BDI class of topological insulators, which are described by an integer $\mbb Z$ topological invariant.\cite{Schnyder2008,Kitaev2009,Ryu2010}

Due to translational invariance, the Hamiltonian can be written in momentum space, $H=\sum_kc^\dag_k\mc H(k)c_k$, with
\begin{eqnarray}
\mc H(k)&=&h_3(k)\sigma_3+h_1(k)\sigma_1,\notag\\
h_1(k)&=&m-g\cos k,\quad h_3(k)=t\sin k,
\end{eqnarray}
where $k$ is the 1D quasimomentum and $\sigma_i$ are the usual spin Pauli matrices. The first term proportional to $\sigma_3$ alone would describe a model with two Dirac points located at $k=0$ and $k=\pi$, respectively. The additional time-reversal-breaking terms, proportional to the Pauli matrix $\sigma_1$, open up gaps at $k=0$, and $k=\pi$, with size $2(m-g)$ and $2(m+g)$, respectively.

For $|m/g|\ne 1$, the system becomes insulating, with a gap $\Delta=2|m-g|$. There is a topological phase transition between two different insulating phases defined by $|m/g|>1$ (trivial phase) and $|m/g|<1$ (topological phase). The two topologically nontrivial phases are distinguished by a topological invariant (the winding number) $w=\sgn(g)$, which is zero in the trivial phase.

\begin{figure}[t]
\includegraphics[width=\columnwidth]{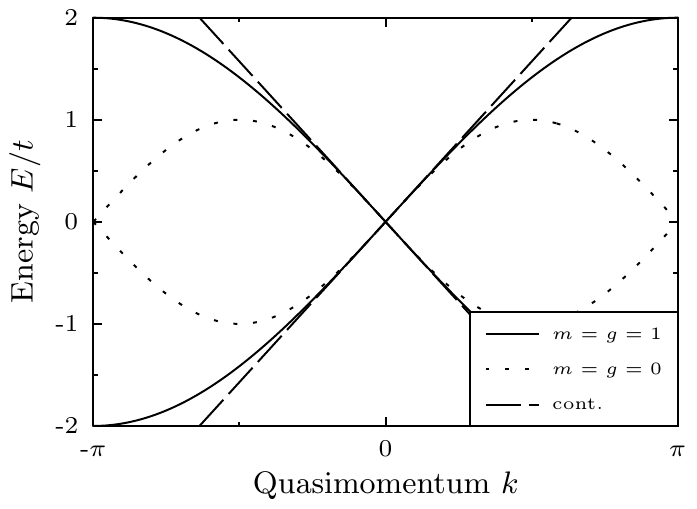}
\caption{Energy dispersion in the clean system, in the absence of magnetic flux. A Dirac point forms at $k=0$, and the spectrum is linear only near the Fermi point. At half filling, the spectrum is particle-hole symmetric around Fermi energy $E_F=0$. By varying the $m=g$ parameter, one can go from a single Dirac cone (straight line) to a double Dirac cone ($m=g=0$) (dotted line). Away from the $k=0$ point, the lattice spectra deviate from the linear dispersion of the continuum model Eq.~(\ref{hamiltonianconti}) (dashed line referred to as ``cont.''). The system parameters are given in units of $t$.}
\label{fig:dispersion}
\end{figure}

We will focus in particular on the phase of the Creutz model where a single massless Dirac fermion coexists with a massive one. For $m=\pm g$ the gap closes at least at one $k$ point. Without loss of generality, let us choose the case $m=g$, where the gap vanishes at momentum $k=0$, while a tunable gap of size $4m$ remains at $k=\pi$ (Fig.~\ref{fig:dispersion}). Then the system can accommodate two flavors of Dirac fermions, one massless at $k=0$ and an additional one at $k=\pi$ with a tunable mass $2m$.

The continuum Dirac Hamiltonian can be seen as an effective model describing the low-energy, long-wavelength limit of the lattice Wilson-Dirac fermion. This will allow us in particular to discuss continuum versus lattice effects. Indeed, the effective Hamiltonian near $k=0$, for a generic $m=g$, reads
\begin{equation}
\mc H(|k|\ll 1)=\hbar v_F k\sigma_3+\frac{mk^2}{2}\sigma_1+O(k^3),
\end{equation}
with the Fermi velocity $v_F=ta/\hbar$. Note that the dispersion of the massless fermion is also affected by the time-reversal-breaking terms at the second order in momentum $k$.

Although the Creutz model breaks time-reversal symmetry, there is still an antiunitary operator $\bar T$, which commutes with the Hamiltonian.
Let us designate the symmetry represented by this operator as a pseudo-time-reversal symmetry (PTRS),
\begin{equation}\label{ptrs}
\bar T\mc H(k)\bar T^{-1}=\mc H(-k),\quad \bar T=\sigma_1\mc K,
\quad\bar T^2=1,
\end{equation}
where $\mc K$ is the complex conjugation operator. At zero flux (or half-integer flux $\Phi/\Phi_0$), all the eigenvalues are degenerate because states with opposite momentum $k$ have the same energy. However, this degeneracy is not protected, if nonmagnetic impurities are added, because the system lacks TRS, and the PTRS operator squares to 1. Moreover, the presence of the PTRS ensures that all nondegenerate states at zero flux will carry zero current (see Sec.~\ref{subsec:scalar}).

\subsection{Persistent current}
The persistent current is defined, at zero temperature, as the derivative of the ground-state energy of the ring, $E(\Phi)$, with respect to the magnetic flux $\Phi$,
\begin{equation}\label{currentDEF}
I(\Phi)=-\frac{\pd E}{\pd\Phi},
\end{equation}
using the sign conventions of Fig.~\ref{fig:AB}: paramagnetic (diamagnetic) persistent current is positive (negative).

At finite temperature $T$ and fixed chemical potential $\mu$, the persistent current is defined as
\begin{equation}\label{currentDEFfinite}
I(\Phi)=-\frac{\pd \Omega}{\pd\Phi},
\end{equation}
where $\Omega(\Phi)$ is the grand potential.

For a noninteracting system, the grand potential is given by
\begin{equation}\label{currentDEFfinite2}
\Omega(\Phi)=-\frac{1}{\beta} \sum_\nu \ln \big[1+e^{-\beta(E_\nu(\Phi)-\mu)}\big],
\end{equation}
with $\beta=1/k_BT$; $k_B$ is the Boltzmann constant, and $T$, the temperature. The index $\nu$ denotes a set of quantum numbers labeling the energy eigenstates of the system. After differentiation with respect to the flux, one obtains the PC
\begin{equation}\label{currentSUMfinite}
I(\Phi)=\sum_\nu f(E_\nu) i_\nu(\Phi),
\end{equation}
where $ i_{\nu}(\Phi)=-\partial E_\nu(\Phi)/\partial \Phi$ is the current carried by a single energy level $E_\nu$, and,
\begin{equation}
f(E_\nu)=\frac{1}{e^{\beta(E_\nu-\mu)}+1},
\end{equation}
is the Fermi-Dirac occupation function. Hence the persistent current is the sum of the currents carried by all single-electron states weighted by the Fermi-Dirac distribution.

Considering the continuum model of Eq.~(\ref{hamiltonianconti}), the single-electron states are labeled by their orbital index $n$, and spin-$\sigma_3$ eigenvalues, $\sigma=\pm$: $\nu = (n,\sigma)$. In the zero-temperature limit, $T \to 0$, the total current is defined as
\begin{equation}\label{currentsum}
I(\Phi)=\sum_{n,\sigma} i_{n\sigma}(\Phi),
\end{equation}
where the sum runs only over occupied states $(n,\sigma)$. In the next section (Sec.~\ref{sec:cleanring}), the problems related to the presence of an infinity of negative energy states in the continuum model from Eq.~(\ref{hamiltonianconti}) will be cured through an ultraviolet regularization. But for the moment, let us focus on the current carried by a single energy level. In the continuum model of Eq.~(\ref{hamiltonianconti}), such a single-state current can be calculated as the derivative,
\begin{equation}\label{defsinglecurrent}
i_{n\sigma}(\Phi)=-\frac{\pd E_n^\sigma}{\pd\Phi}=-\sigma \frac{e v_F}{L},
\end{equation}
where $L=2 \pi R$ is the total length of the ring and $\sigma=\pm 1$ are the eigenvalues of spin $\sigma_3$ matrix. Note that the current does not depend explicitly on the orbital index, and all spin-up (respectively spin-down) states carry the same diamagnetic (respectively paramagnetic) current.

This result can also be derived from the current operator for a single electron,
\begin{equation}\label{defsinglecurrentoperator}
i=-\frac {e v_F}{L} \sigma_3,
\end{equation}
whose quantum average
\begin{equation}\label{defcurrent2}
\la n,\sigma|i|n',\sigma'\ra=-\sigma\frac{e v_F}{L}\delta_{nn'} \delta_{\sigma\sigma'}=i_{n\sigma}(\Phi),
\end{equation}
is diagonal in the basis of the energy eigenstates.

Finally, in the lattice model, the current operator reads
\begin{equation}\label{currentOp}
\mc J=-\frac{\pd H_{\rm Creutz}}{\pd\Phi}=
\frac{I_0}{2t}\sum_jc^\dag_j(t\sigma_3+ig\sigma_1)
e^{i\phi}c_{j-1}+\hc,
\end{equation}
where the sum runs over all sites $j$ in the ring. As already noticed, $I_0=ev_F/L$ is the absolute value of the current carried by one eigenstate, and the Fermi velocity in the tight-binding model is $v_F=ta/\hbar$.

The PC is obtained again from a sum of the currents carried by all occupied energy eigenstates $|n\ra$ of the Hamiltonian
\begin{equation}
I(\Phi)
%=-\sum_{\text{occ.} n}\big\la n\big|\frac{\pd H}{\pd\Phi}\big|n\big\ra
=\sum_{\text{occ. } n}\la n|\mc Jn\ra.
\end{equation}

\section{Dirac fermions in a clean ring}
\label{sec:cleanring}
%\begin{figure}[t]
%\centering
%\includegraphics[width=4.5cm]{AB}
%\caption{The helical model~(\ref{hamiltonianconti}) is implemented on a ring of length $L=2 \pi R$, threaded by a tube of magnetic flux $\Phi$. There are two counter-propagating spin $\sigma_3$-eigenstates denoted by $\up$, and $\down$ arrows. 
%The arrow in the azimuthal direction indicates the convention for positive (paramagnetic) persistent current $I(\Phi)$.
%}
%\label{fig:AB}
%\end{figure}
In this section, we consider the persistent current flowing in a perfectly clean Dirac ring at zero temperature, using the two models introduced in previous section. For the continuum model (also describing the helical edge state of a quantum spin Hall droplet), it is necessary to use a regularization procedure to extract the PC carried by the infinite Fermi-Dirac sea. In the lattice ring, the spectrum is automatically bounded, and the total energy can be computed directly, without any regularization procedure. It is shown that the two models lead to the same persistent current, $I(\Phi)$, over a wide range of parameters, provided the low-energy spectra coincide. This was not \textit{a priori} trivial since on one side, the PC is a thermodynamical observable, which depends on the whole spectrum, and, on the other side, the two models have very different high-energy states.

\subsection{Continuum helical model}
\label{subsec:clean}
It is well known that the Dirac Hamiltonian suffers from having an infinite number of negative energy states. Historically, Dirac solved this problem by supposing that all the negative states are filled, forming the ``Dirac sea,'' and proposed the hole picture, which led to the prediction of antiparticles.\cite{Dirac1928,*Dirac1930} In the present context of condensed matter physics, the negative energy states are the occupied states of the valence band and the ``antiparticle'' states correspond to hole quasiparticles.
The spectrum is bounded, the Dirac Hamiltonian being a low-energy approximation near the Dirac point. Nevertheless it would be interesting to extract the exact persistent current using only the linearized low-energy effective Dirac Hamiltonian. This can be done using an ultraviolet regularization, where the energy states deep in the Dirac sea have an exponentially small contribution to the physical properties of the system. The next section follows closely Refs.~\onlinecite{Manton1985, Shifman1991} in obtaining a regularized total energy.

Let us perform here a grand-canonical calculation with the Fermi energy fixed at $E_F=0$. One defines the total regularized charges for right- $(+)$ and left- ($-$) movers as
\begin{equation}\label{charges}
Q^\pm=\sum_{n=\mp\infty}^{N^\pm_\varphi}e^{\e E_n^\pm/\hbar\omega},
\end{equation}
where $N^{+}_\varphi$ ($N^{-}_\varphi$) is the index of the highest occupied energy level for right-moving, spin-up state (left-moving, spin-down state). The constant $\e$ is an infinitesimally small, positive, real number. Note that for $\e=0$, $Q^\pm$ counts the (infinite) number of spin-up $(+)$ and, respectively, spin-down $(-)$ occupied energy states. For finite $\e$, these series are made convergent, because energy states below the cutoff $\Lambda_\e =-\hbar \omega/\e$ are exponentially suppressed. 

\begin{figure}[t]
\centering
\includegraphics[width=0.7\columnwidth]{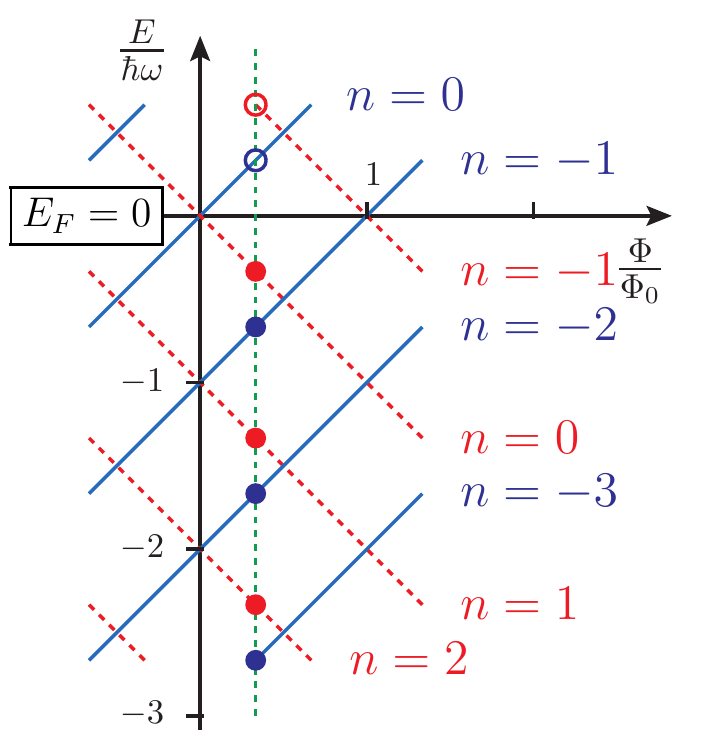}
\caption{(Color online). Energy-flux characteristic $E^\pm_n=\pm\hbar\omega(n+\Phi/\Phi_0)$. Fermi energy $E_F$ is fixed at zero, and the flux $\Phi/\Phi_0$ takes an arbitrary value (represented by the dashed green line) in the interval $(0,1)$. Filled (empty) spin states are represented by $\bullet$ ($\circ$). The spin-up (down) states' evolution under flux is represented by dashed red (solid blue) lines. In the flux interval $(0,1)$, the infinite occupied number of spin-up states are indexed by integer $n$, $n\le -1$, while occupied spin-down states, by $n\ge 0$.}
\label{fig:energyflux}
\end{figure}

\begin{figure}[t]
\centering
\includegraphics[width=\columnwidth]{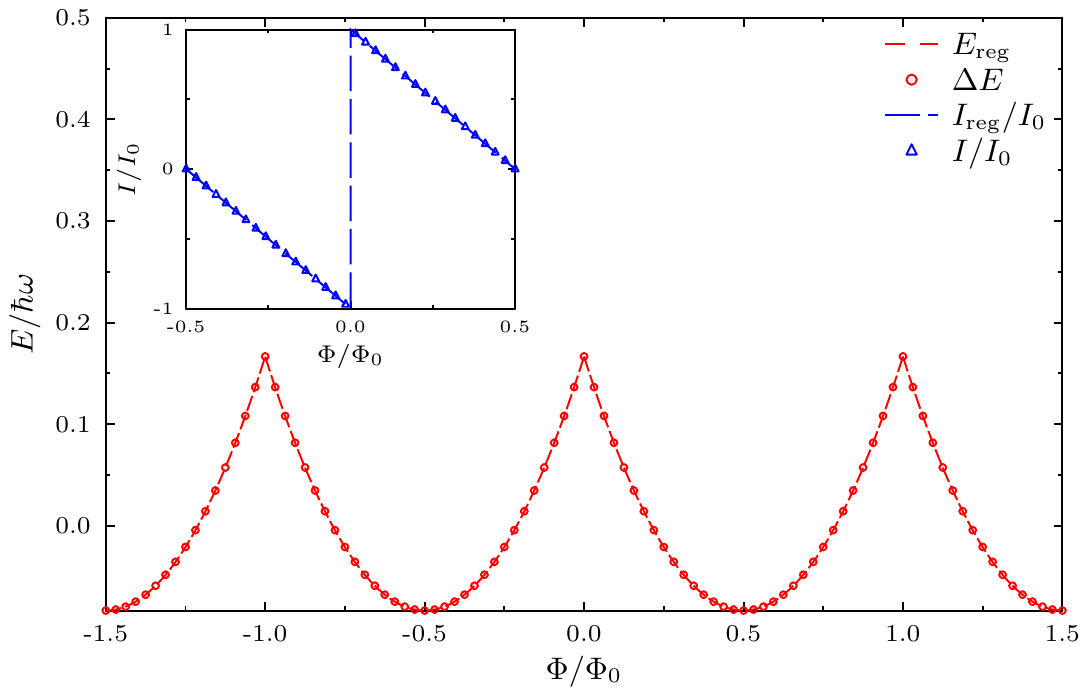}
\caption{(Color online) The regularized ground-state energy of the ring as a function of the magnetic flux, $\Phi/\Phi_0$, from Eq.~(\ref{totalenergy}).
The regularized energy follows perfectly the variation in the total energy obtained numerically in the lattice model~(\ref{creutz}); $\Delta E(\Phi)=E(\Phi)-E_{\rm min}-\hbar\omega/12$ (in units of $\hbar\omega$) is represented by red open circles.
In the inset is represented the dimensionless persistent current $I/I_0$ as a function of flux. The analytical result matches exactly the numerical calculation. The parameters are $\hbar\omega=hv_F/L$ and $I_0=ev_F/L$, where $v_F$ is the Fermi velocity and $L=300$ the circumference of the wire.}
\label{fig:effenergy}
\end{figure}

Owing to the flux periodicity, it is sufficient to consider the reduced flux, $\Phi/\Phi_0$, in the interval between $0$ and $1$. As shown in Fig.~\ref{fig:energyflux}, the occupied spin-up states are labeled by orbital indices ranging from $n=-\infty$ to $n=N^+_\varphi=-1$, while the occupied spin-down states are labeled from $n=N^-_\varphi=0$ to $n=\infty$. As a result, the geometric sums in Eq.~(\ref{charges}) read
\begin{equation}\label{chargeplusexact}
Q^+=\sum_{n=-\infty}^{-1}e^{\e E_n^+/\hbar\omega}=\frac{e^{\e(\Phi/\Phi_0-1)}}{1-e^{-\e}},
\end{equation}
and 
\begin{equation}\label{chargeminusexact}
Q^-=\sum_{n=0}^{\infty}e^{\e E_n^-/\hbar\omega}=\frac{e^{-\e\Phi/\Phi_0}}{1-e^{-\e}}.
\end{equation}
The power expansion in small parameter $\e$ reads
\begin{equation}\label{chargeplusmoins}
Q^\pm=\frac{1}{\e}\pm\bigg(\frac{\Phi}{\Phi_0}-\frac{1}{2}\bigg)+\frac{\e}{2}\bigg(\frac{\Phi^2}{\Phi_0^2}-\frac{\Phi}{\Phi_0}+\frac{1}{6}\bigg)+O(\e^2).
\end{equation}
This procedure has singled out an infinite contribution ($1/\e$) to the charges from a finite and flux-dependent contribution.
The next step in the regularization procedure aims to cancel the flux-independent infinities. From a physical point of view, the total electric charge of ground state must be zero. This leads us to introduce the total regularized charge $Q_{\rm reg}=Q^+ + Q^- - 2/\e$, which reads
\begin{equation}\label{chargeregul}
Q_{\rm reg}=\e \bigg(\frac{\Phi^2}{\Phi_0^2}-\frac{\Phi}{\Phi_0}+\frac{1}{6} \bigg)+O(\e^2).
\end{equation}
Note that in the limit $\e\to 0$, one recovers the expected physical result: $Q_{\rm reg}=0$.
The total regularized energy, $E(\Phi)$, is obtained from the formal sums in Eq.~(\ref{charges}),
\begin{equation}\label{toterg}
E(\Phi)=\lim_{\e\to 0}\hbar\omega\frac{\pd Q_{\rm reg}}{\pd\e}.
\end{equation}
Therefore the total energy in the flux interval $\Phi/\Phi_0\in(0,1)$ explicitly reads
\begin{equation}\label{parabola}
E(\Phi)=\hbar\omega \bigg[\bigg(\frac{\Phi}{\Phi_0}-\frac{1}{2}\bigg)^2-\frac{1}{12}\bigg].
\end{equation}
Owing to the $\Phi_0$ periodicity, this result can be immediately extended to arbitrary noninteger flux $\Phi/\Phi_0$:
\begin{equation}\label{totalenergy}
E(\Phi)=\hbar\omega \bigg[\bigg(\frac{\Phi}{\Phi_0}-\frac{1}{2}
-\bigg\lfloor
\frac{\Phi}{\Phi_0}-\frac{1}{2}\bigg\rceil\bigg)^2-\frac{1}{12}\bigg],
\end{equation}
where $\lfloor x \rceil$ denotes the rounding of real $x$ to the nearest integer (see Fig.~\ref{fig:effenergy}).

The total energy can be Fourier analyzed as
\begin{equation}
\label{totalenergyFourier}
E(\Phi)=\frac{\hbar \omega}{\pi^2}\sum_{m=1}^{\infty}
\frac{1}{m^2}\cos\bigg(2\pi m\frac{\Phi}{\Phi_0}\bigg).
\end{equation}
The flux-independent constant $\hbar\omega/12$ could have been easily thrown away in the regularization procedure together with the infinite flux-independent term $1/\e$. It was conveniently kept to cancel the flux-independent constant, $m=0$ term, in the Fourier expansion for the energy.

Even if the spin-up and spin-down electrons have a dispersion linear in the magnetic flux, the regularization has determined a quadratic total energy between any two integer values of the flux $\Phi/\Phi_0$. Also notice that the final result is a $\Phi_0$-periodic function in magnetic flux, reminiscent of the total energy in the case for nonrelativistic fermions.\cite{Cheung1988, Cheung1989}

The persistent current at zero temperature is determined from the energy-flux characteristic from Eqs.~(\ref{totalenergy}) and (\ref{totalenergyFourier}). The PC-flux relation has a sawtooth shape (inset of Fig. \ref{fig:effenergy}), which is identical to the $I(\Phi)$ curve for nonrelativistic fermions.\cite{Byers1961,Cheung1988} The current reads
\begin{eqnarray}\label{currClean}
I(\Phi)&=&-\frac{\pd E}{\pd\Phi}=\frac{2I_0}{\pi}
\sum_{m=1}^{\infty}
\frac{1}{m}\sin\big(2\pi m\frac{\Phi}{\Phi_0}\big),\notag\\
\frac{I(\Phi)}{2I_0}&=&\frac{1}{2}-\frac{\Phi}{\Phi_0}
-\bigg\lfloor\frac{1}{2}-\frac{\Phi}{\Phi_0}\bigg\rceil,
\end{eqnarray}
where $I_0=ev_F/L$ is the maximal current carried by one spin eigenenergy.

The current presents discontinuities (cusps in the total energy) at integer reduced fluxes $\Phi/\Phi_0$. This can be understood by noticing that the states at $E=0$ are doubly degenerate. Upon an infinitesimal increase of the flux, the spin-down state becomes occupied and the spin-up state becomes empty. In contrast, an infinitesimal decrease of the flux reverses the situation. Hence increasing the flux infinitesimally from $\Phi/\Phi_0=0$ has a net effect of replacing a spin-up by a spin-down state, thereby yielding a positive (paramagnetic) jump of $2 I_0$ in the current, because a spin-up (respectively spin-down) level carries a current $-I_0$ (respectively $+I_0$).  

\subsection{Creutz lattice model}
For electrons on a lattice, the fact that the Brillouin zone (BZ) is a compact manifold forces the linear dispersion to bend in such a way to satisfy periodicity in the BZ. This generates automatically a finite bandwidth and the regularization is automatic. The presence of a lattice also allows us to count electrons in occupied levels and therefore to discuss even/odd parity effects. Here we provide for the reader who might (legitimately) feel suspicious about the regularization procedure a comparison between the regularization result and the well-defined total energy of the lattice system.

\subsubsection{Continuum-lattice comparison}
The regularized model only captures the variation in the total energy of the ring as a function of the flux.
But the total energy in the regularized model Eq.~(\ref{totalenergy}) cannot be directly compared with the results in the lattice. 
The analytical result gives the variation of the total energy and not a constant, flux-independent, background energy. Thus it manages to 
correctly predict the persistent charge currents. The regularized energies and currents reproduce perfectly the numerical result (see Figs.~\ref{fig:effenergy} and~\ref{fig:parity}). From simulations, the variation in the total energy $\Delta E(\Phi)$ is defined as
\begin{equation}
\Delta E(\Phi)=E(\Phi)-E_{\min}-\hbar\omega/12,
\end{equation}
where $E(\Phi)$ is the total energy and $E_{\min}$ is the minimum value of total energy as a function of flux. In the present model, the energy is minimum at half-integer flux $\Phi/\Phi_0$. The comparison with the regularized result requires finally the subtraction of the flux-independent constant $-\hbar\omega/12$, which is of no consequence for the PC.

The essential feature of the total energy is its parabolic flux dependence between two integer fluxes $\Phi/\Phi_0$. This explains why the PC in the Dirac clean system has a sawtooth flux dependence as in the more familiar cases\cite{Imry2008, AkkM2011} of nonrelativistic fermions with quadratic dispersion (inset of Fig.~\ref{fig:effenergy}).

\subsubsection{Parity effects}
Until now, the focus has been on the models at half filling. It is interesting to discuss parity effects due to removal (addition) of an even/odd number of electrons from the system. From a comparison between the lattice total energy and the regularized total energy, it is possible to infer that
the half-filled case $(\mu=0)$ corresponds to a band with an even number of states filled.

In the continuum system such effects are readily understood. For example, removing an electron of any spin decreases the total energy by $\hbar\omega/2$. Let us study again the evolution of level in the magnetic flux, in Fig.~\ref{fig:energyflux}, with $\Phi/\Phi_0\in(0,1)$.
The removal of one electron moves the Fermi energy to $\mu=-\hbar\omega/2$. Consequently the spin-degenerate Fermi energy states are placed at half-integer flux instead of integer flux.
Relabeling the states allows one to recover the half-filled case, with the essential change that the energy- and PC-flux characteristics have been shifted by half-flux quantum.

Owing to spectrum periodicity as a function of the flux, a removal or addition of an even number of electrons recovers the half-filled case. In contrast, a removal or addition of an odd number of electrons from half filling amounts to a shift of the flux by half-flux quantum. In terms of wave functions, this behavior is equivalent to changing from periodic boundary conditions to antiperiodic boundary conditions.

\begin{figure}[t]
\includegraphics[width=\columnwidth]{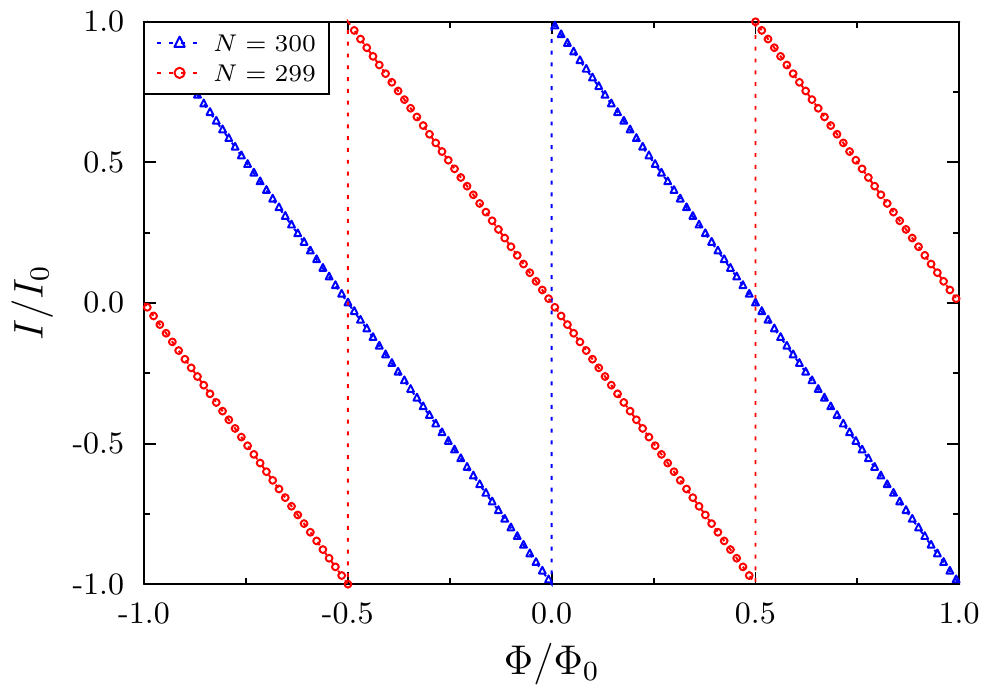}
\caption{(Color online) Parity effects due to the variation in the number of fermions from half filling. For a $L=300$ site ring, removing an odd number of fermions fermion has the effect to produce a shift in the PC-flux characteristic. Compare the half-filling PC, occupied states $N=300$ (represented by blue triangles), with an odd number of fermions, $N=299$ (in red circles).}
\label{fig:parity}
\end{figure}

Parity effects are equally present in the lattice. Changing the chemical potential by removing one electron produces the predicted shift in the PC-flux characteristic (Fig.~\ref{fig:parity}).

\subsubsection{Crossover from one to two Dirac points}
\label{subsec:cross}
A remarkable feature of the model is the presence of two Dirac fermions for a particular choice of parameters $(m=g=0)$. In this case there will be an equal contribution to the PC from both cones. This leads to a doubling in the amplitude of the current.

By increasing $m=g$, one of the Dirac points becomes gapped. The contribution to the PC from this gapped Dirac branch is exponentially suppressed.\cite{Kohno1992} The amplitude will decrease with the system size and the amplitude of the gap.

\begin{figure}[t]
\includegraphics[width=\columnwidth]{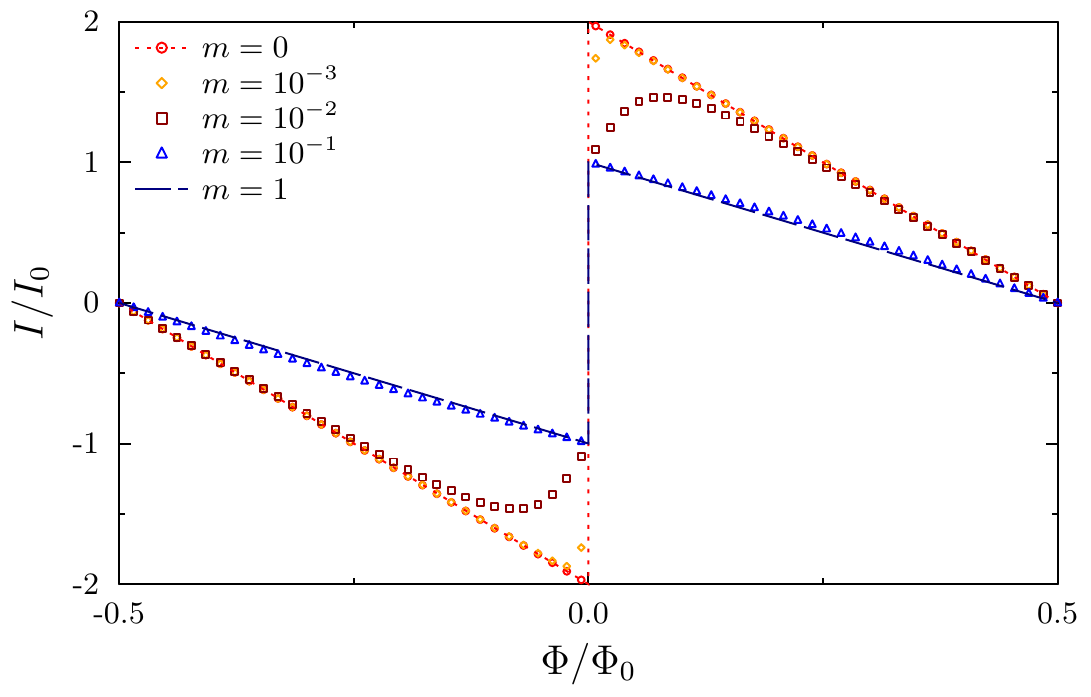}
\caption{(Color online) Current-flux characteristic in the crossover region, from two Dirac cones to single Dirac cone. When removing the cone at $k=\pi$, with a gap $4m$, the PC drops suddenly to the single Dirac cone PC for $m\sim 1/L$. The crossover is seen more easily in small systems (here $L=20$ sites).}
\label{fig:cleanCross}
\end{figure}

In the present case, let us gap the cone at $k=\pi$ (see Fig.~\ref{fig:dispersion}). The gap at this point reads $\Delta_\pi=2|m+g|$. On the lattice, any effects from the crossover will be seen at a scale where $\Delta_\pi\sim 1/L$. Therefore, if the gap at $k=\pi$ becomes larger than $1/L$, the maximal amplitude of the PC will seem to jump directly from $2I_0$ in the two Dirac case to $I_0$ for a single Dirac cone. This effect is very difficult to observe for large rings, because as soon as $m=g$ is finite, a tiny gap is opened and the current amplitude is immediately halved. A crossover between the two cases due to this subtle finite-size effect is presented in Fig.~\ref{fig:cleanCross} for a small ring $L=20a$. In the crossover region, where the current interpolates between the curves for one and two Dirac points, the gapped Dirac points still carries a finite contribution to the PC. 

\subsubsection{Comparison with Josephson junctions}
In this paper, we investigate the persistent current in purely normal (nonsuperconducting) rings pierced by a magnetic flux. Nevertheless, we would like to stress here a specific analogy between the PC in normal rings and the DC supercurrent in Josephson junctions (JJs). Let us consider a JJ consisting of a single narrow ballistic channel (of length $d_N$) contacted between two superconducting electrodes, whose phases differ by $\chi$. Below the gap, ($\epsilon < \Delta_0$), the discrete phase-dependent Andreev levels $\epsilon_n(\chi)$ are trapped in the normal region, while extended excitations form a continuum above the gap. In long junctions, i.e., for $d_N \gg \hbar v_F/\Delta_0 $, the low-energy Andreev levels ($\epsilon_n(\chi) \ll \Delta_0$) and the total supercurrent $I_J(\chi)$ are piecewise linear in the phase $\chi$. \cite{Bardeen:1972,Affleck:2000} Moreover, the total DC supercurrent, which is carried by all the phase-dependent energy levels (below and above $\Delta_0$) can be evaluated from the knowledge of the zero-energy (Fermi-level) Andreev scattering.\cite{Cayssol:2003} In particular, this implies that the sawtooth shape of $I_J(\chi)$ is completely insensitive to the details of the high-energy level flux dependence.

Therefore the Andreev spectrum of long SNS junctions is very similar to the flux-dependent spectrum of the electronic states in the normal ring (Fig.~\ref{fig:energyflux}). In the continuum model, the electronic levels evolve linearly in flux at all energies. This is similar to the limit $\Delta/\epsilon \to \infty$ in the SNS junction case. In the Creutz lattice model, we have checked that the sawtooth current-flux relation holds, and it is insensitive to the details of the high-energy spectrum (where the levels no longer evolve linearly in flux). Nevertheless in Sec.~\ref{subsec:cross} (Fig.~\ref{fig:cleanCross}), we have followed the current-phase relation of the normal ring when the band structure is continuously modified from a single to two Dirac points. Apart from the trivial factor 2, there are indeed observable deviations from the sawtooth $I(\phi)$, but only when the spectrum is nonlinear \textit{near} the Fermi level, which would correspond, in the SNS junction analogy, to the short junction limit.

\section{Ring with a single impurity}
\label{sec:singleimpurity}
This section investigates the effect of a single impurity on the persistent current flowing in Dirac rings. First is discussed the case of a spin-independent impurity potential (scalar impurity) proportional to the identity matrix in spin space $\sigma_0$ (Sec. \ref{subsec:scalar}). Second, we consider a magnetic impurity that flips the spin via a potential proportional to the $\sigma_1$ matrix (Sec. \ref{subsec:magnetic}). A magnetic impurity, that breaks time-reversal symmetry, is in general more harmful to the persistent current than a scalar nonmagnetic impurity. For both types of impurity, we compare in detail the helical continuum model and the lattice model.  For the helical continuum model, the current-flux relation $I(\Phi)$ is obtained analytically, and its agreement with the lattice model is discussed. 

\subsection{Scalar impurity }
\label{subsec:scalar}
Here we consider a ring with a single scalar impurity acting through a potential proportional to the identity matrix in spin space $\sigma_0$. We show that such perturbation has rather different effects in the continuum helical model and in the lattice model.     

\subsubsection{Continuum helical model}

Let us examine the Hamiltonian in Eq.~(\ref{hamiltonianconti}), supplemented by a potential term, $U(\theta) \sigma_0$, which affects identically the spin-up and spin-down states:
\begin{equation}
\label{hamiltonianImpurityScalar}
H=\hbar \omega(-i\pd_\theta+\Phi/\Phi_0)\sigma_3+U(\theta)\sigma_0.
\end{equation}
This Hamiltonian remains spin-diagonal, which means that the spin-up and spin-down channels are still decoupled in the presence of the impurity. The impurity potential only produces phase shifts for right-moving (spin-up) and for left-moving (spin-down) carriers without inducing backscattering between the two types of chiral particles. Therefore components $\psi_+$ and $\psi_-$ of the spinor wave function obey decoupled equations:  
\begin{equation}\label{decoupspin}
\bigg[\pm\hbar\omega\bigg(-i\pd_\theta+\frac{\Phi}{\Phi_0}\bigg)
+U(\theta)\bigg]\psi_\pm=E\psi_\pm.
\end{equation} 
To be more definite, let us choose a sharp barrier model for the potential:
\begin{equation}\label{potential}
U(\theta)=
\begin{cases}
U,&\theta\in[0,\alpha), \\
0,&\theta\in[\alpha,2\pi),
\end{cases}
\end{equation}
where $\alpha$ fixes the angular extension of the potential.
By imposing periodic boundary conditions, $\psi_\pm(2\pi)=\psi_\pm(0)$, one finds the energy eigenvalues
\begin{equation}\label{energyDecoup}
E_n^\pm=\pm \hbar\omega(n+\Phi/\Phi_0)+\frac{U \alpha}{2\pi}.
\end{equation}

Let us now consider the limiting case of a delta-function potential, namely $U(\theta)=U_0\delta(\theta)$, which corresponds to $U \to \infty$ and $\alpha \to 0$, while keeping $U \alpha$ fixed. This limit must be treated with care, because the equations for chiral fermions Eq.~(\ref{decoupspin}) may yield a different scattering phase depending on the regularization scheme.\cite{Gogolin2004} Both the spectrum energy cutoff, $\Lambda_\epsilon$, and the impurity barrier height, $\sim 1/\alpha$, are sent to infinity, and the result depends on which limit is taken first. For the helical continuum model, we work in the infinite-bandwidth approximation: the energy bandwidth is implicitly infinite, and the limit $1/\alpha\to\infty$ is taken last. 

In this infinite-bandwidth approximation, the spectrum of the continuum model ring with a delta scatterer is
\begin{equation}\label{energyDecoupDelta}
E_n^\pm=\pm \hbar\omega(n+\Phi/\Phi_0)+\frac{U_0}{2\pi},
\end{equation}
where $U(\theta)= U_0 \delta(\theta)$. The potential strength $U_0$ does not depend on the flux and enters as an additive quantity to the energy. As can be seen from Eq.~(\ref{energyDecoupDelta}), the scalar impurity is just shifting the spectrum by a global constant with respect to the clean-ring spectrum from Eq.~(\ref{spectrumconti}). In particular, at external fluxes $\Phi$, proportional to an integer (or half-integer) multiple of the flux quantum $\Phi_0$, the energy level crossings of the clean spectrum (Fig.~\ref{fig:energyflux}) are preserved. This is a manifestation of the absence of backscattering in the helical liquid described by the model in Eq.~(\ref{hamiltonianconti}). The helical model in Eq.~(\ref{hamiltonianconti}) is time-reversal invariant at these values of the flux, and the scalar impurity does not break TRS.    

In conclusion, within the helical model, the persistent current-flux relation $I(\Phi)$ is unaffected by a scalar (nonmagnetic) impurity, $U_0 \delta(\theta)$. All the energy levels are simply shifted by a common energy offset, $\frac{U_0}{2\pi}$.

\subsubsection{Creutz lattice model}
Now let us turn to the Creutz lattice model in presence of a single scalar impurity. The scalar scatterer is modeled by adding to Eq.~(\ref{creutz}) an on-site spin-independent energy  $U_0^s$, 
\begin{equation}
H_s=H_{\rm Creutz}+U_0^sc^\dag_J\sigma_0c_J,
\end{equation}
located at an arbitrary site $J$.

If the impurity strength is small with respect to the bandwidth, the mapping between the lattice impurity strength and the continuum model has the simple form
\begin{equation}
\frac{U_0}{\hbar\omega}=\frac{U_0^s}{t}.
\end{equation}
In the continuum model, the impurity strength $U_0$ is expressed in units of energy level spacing at the Dirac point, $\hbar\omega$, and, on the lattice, $U_0^s$, in units of hopping strength $t$. For large potential strength, one needs to renormalize the potential in the continuum model (Sec.~\ref{subsubsec:CrMag}).

\begin{figure}
\includegraphics[width=0.75\columnwidth]{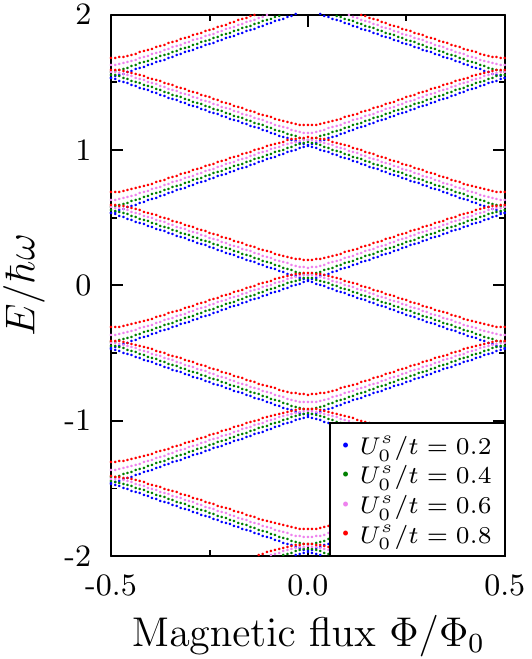}
\caption{(Color online) Energy-flux dependence in the Creutz model~(\ref{creutz}) with a single scalar potential $U_0^s$. In contrast to the continuum results, the time-reversal-symmetry breaking in the lattice opens up gaps in the spectrum at (half-)integer reduced flux $\Phi/\Phi_0$. The model parameters are $L=300a$ and $m=g=t$ in units where $t=1$.}
\label{fig:deltaScal}
\end{figure}

Several spectra, obtained from the numerical diagonalization of $H_s$, are shown in Fig.~\ref{fig:deltaScal} for different scalar impurity strengths. The striking feature is the opening of gaps at integer and half-integer fluxes, $\Phi/\Phi_0$ (see Fig.~\ref{fig:deltaScal}). This is at odds with the above result in the continuum helical model, in which a scalar impurity was unable to remove the Kramers degeneracy between levels at these fluxes. 

The explanation is that the helical model respects time-reversal invariance represented by $T$ (with $T^2 = -1$) at those fluxes, whereas the Creutz model breaks TRS, and has only a pseudo-time-reversal invariance, $\bar T$ (with $\bar{T}^2=1$). The microscopic spin-mixing terms of the Creutz model break time-reversal symmetry, $T$ (even when the external flux is a multiple of the flux quantum $\Phi_0$), and spoil the Kramers protection, even against scalar disorder.

Nevertheless, the presence of the pseudo-time-reversal symmetry (PTRS) has some consequences at (half-)in\-teger flux.\cite{Kravtsov1992} Indeed, if $|n\ra$ is a nondegenerate energy eigenstate, then one can define the real eigenenergy state $|n'\ra=|n\ra+\bar T|n\ra$ as a $\bar T$ eigenstate with $+1$ eigenvalue. Therefore
\begin{equation}
\la n'|\mc J n'\ra=\la n'\bar T|\mc J\bar T n'\ra=-\la n' |\mc J n'\ra=0,
\end{equation}
where the second equality follows from the fact that the current operator $\mc J$ from Eq.~(\ref{currentOp}) anticommutes with the PTRS operator at (half-)integer flux. Finally, each nondegenerate individual level carries a vanishing current at $\Phi=n\Phi_0/2$ (with $n$ an arbitrary integer). This results in a smoothing of the clean case discontinuities in the $I(\Phi)$ curve, even at $T=0$, and in a decrease of the maximal PC. 

The only eigenstates that may carry current at zero flux are the degenerate states.
From the numerical simulations and analytical approximations at $\Phi=0$, it follows that a doublet of degenerate states exists at energy
\begin{equation}
E_d=\frac{U^s_0t}{m}.
\end{equation}
This energy corresponds to a resonant state in the band. Note that this energy is usually rather high in terms of typical level spacings $\hbar \omega$. For instance, with the parameters of Fig.~\ref{fig:deltaScal}, namely $t=m$, this degenerate doublet would appear at $E_d/\hbar \omega=U^s_0/\hbar \omega=(U^s_0/t)\cdot(R/a)$, where the factor $R/a$ is usually large for realistic rings.  

In conclusion, a scalar impurity can bring a sizable decrease in the persistent current in the Creutz lattice model in contrast to the helical continuum model. However, this decrease remains smaller than the decrease induced by a magnetic impurity as described below. 

\subsection{Magnetic impurity}
\label{subsec:magnetic}

Here we consider a ring with a single magnetic impurity acting as a local Zeeman coupling proportional to the Pauli matrix $\sigma_1$. Since time-reversal $T$ flips all the components of the spin, a potential proportional to any Pauli matrix will break time-reversal symmetry. Nevertheless, for the helical model, a potential proportional to $\sigma_3$ is expected to be less harmful to the PC than potentials proportional to $\sigma_1$ or $\sigma_2$. This is because $\sigma_3$ defines the quantization axis of the helical liquid, and Zeeman perturbations along this axis leave the system gapless.\cite{Ilan2012}

\subsubsection{Continuum helical model}

{\it Ring spectrum.} Since it breaks TRS, a magnetic potential can lead to backscattering processes and to the opening of gaps in the flux-dependent energy spectrum. This leads to a suppression of the persistent current with respect to the clean case.
Let us consider the Hamiltonian Eq.~(\ref{hamiltonianconti}), plus a spin-mixing barrier $U(\theta) \sigma_1$,
\begin{equation}\label{hamOneMag}
H=\hbar \omega(-i\pd_\theta+\Phi/\Phi_0)\sigma_3+U(\theta)\sigma_1,
\end{equation}
where the potential $U(\theta)$ has the rectangular shape defined in Eq.~(\ref{potential}).

The eigenvalue problem $H\Psi(\theta)=E\Psi(\theta)$ for the wave function $\Psi(\theta)$ reads
\begin{equation}
\frac{\partial \Psi}{\partial \theta}=\bigg(\frac{iE\sigma_3+U\sigma_2}{\hbar\omega}-i \frac{\Phi}{\Phi_0}\sigma_0\bigg)\Psi,
\end{equation}
which can be solved in the transfer matrix formalism. The transfer matrices read respectively 
\begin{equation}
T(0,\alpha)=\exp\bigg[\bigg(\frac{i E\sigma_3+U\sigma_2}{\hbar\omega}-i\frac{\Phi}{\Phi_0}\sigma_0\bigg)\alpha\bigg],
\end{equation}
in the region with a finite potential, and, 
\begin{equation}
T(\alpha,2\pi)=
\exp\bigg[i\bigg(\frac{E}{\hbar\omega}\sigma_3-\frac{\Phi}{\Phi_0}\sigma_0\bigg)(2\pi-\alpha)\bigg],
\end{equation}
in the region of vanishing potential, $U(\theta)=0$.
Then the energy quantization condition follows from the fact that the wave function comes back to itself after a circuit around the ring:
\begin{equation}
\det\big[T(\alpha,2\pi)T(0,\alpha)-1]=0.
\end{equation}
This secular equation determines the eigenenergies in the system. Let us further simplify the problem by considering a Dirac-delta magnetic potential, $U(\theta)=U_0\delta(\theta)$. The strength of the delta potential is denoted by $U_0$. This change implies that the angular the angular width becomes infinitesimally small, $\alpha\to 0$, while $\int d\theta U(\theta)= U_0$.
It is important to note that in this limiting procedure, we are implicitly working in the infinite-bandwidth approximation, which is appropriate for potentials smaller or on the order of $\hbar\omega$. As the impurity strength is ramped up to infinity, the transfer matrix over the impurity becomes ill-defined.\cite{Gogolin2004} In the infinite-bandwidth approximation, the quantization condition is determined by the equation
\begin{equation}
\cos\bigg(\frac{2\pi E}{\hbar\omega}\bigg)
\cosh\bigg(\frac{U_0}{\hbar\omega}\bigg)=\cos\bigg(2\pi\frac{\Phi}{\Phi_0}\bigg).
\end{equation}

\begin{figure}[t]
\centering
\includegraphics[width=0.75\columnwidth]{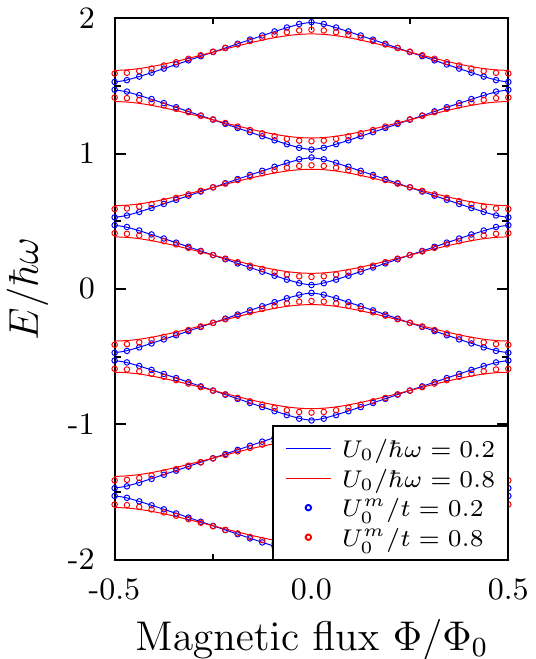}
\caption{(Color online) Energy levels as a function of the magnetic flux for a single magnetic Dirac-delta impurity potential of strength $U_0^m$. The straight lines correspond to the analytical result~(\ref{magenergy}), while the markers correspond to exact diagonalization of the Creutz model~(\ref{creutz}) for an $L=300a$ system. The analytical result matches the numerics for small impurity strength. The two results begin to deviate when $U_0^m$ is on the order of $\hbar\omega$.
Increasing the magnetic potential $U_0^m$ leads to a flattening of the bands.
On the lattice, $\hbar\omega=2\pi ta/L$, $m=g=t$ in units where $t=1$ and lattice spacing $a=1$.}
\label{fig:deltaImpGaps}
\end{figure}

Then it follows from the quantization condition that the band energies read
\begin{equation}\label{magenergy}
E^\pm_n=\hbar \omega \big(n \pm \varphi(\Phi)\big),
\end{equation}
where $n$ is an integer and $\varphi(\Phi)$ is an effective phase defined by
\begin{equation}\label{effectivephase}
\varphi(\Phi)=\frac{1}{2\pi}
\arccos\bigg[\frac{\cos(2\pi\Phi/\Phi_0)}{\cosh(U_0/\hbar\omega)}\bigg].
\end{equation}
Each individual level is $\Phi_0$-periodic in flux and it is labeled by a pair of indices $(n,\lambda=\pm)$ ($n$ an integer). In contrast, the levels in the clean ring were linear in flux Eq.~(\ref{spectrumconti}). With the addition of the magnetic potential, the $\sigma_3$ spin is no longer a good quantum number. 
Consequently, the index $\lambda$ for the eigenenergies no longer describes spin-up and down states.

In the clean systems, there were spin-degenerate states at integer and half-integer flux. In contrast, the magnetic impurity couples the spins and opens up gaps at these values. Figure~\ref{fig:deltaImpGaps} illustrates the energy-flux dependence from Eq.~(\ref{magenergy}) at some small magnetic impurity potentials. Increasing the strength of the impurity potential leads to an increase of the gaps, and to a flattening of the energy levels as a function of flux. Therefore the current carried by each level decreases, and consequently the overall persistent current is also expected to be suppressed.

{\it Persistent current.} In order to obtain a quantitative expression for the reduction to the PC within the helical model, one needs to take into account the contribution from an infinite number of states. As for the clean ring (Sec. \ref{subsec:clean}), this requires us to use a regularization scheme to extract physical information. Here we use a gauge-invariant regularization of the current itself. \cite{Kohno1992} The PC at half filling is determined by adding all the individual currents carried by levels below the Fermi surface:  
\begin{equation}\label{regCurr}
I\big|_{\mu=0}
=\lim_{\epsilon\to0}\bigg(\sum_{n=-\infty}^{n=-1}
i_n^+e^{\e E_n^+/\hbar\omega}
+\sum_{n=-\infty}^{n=0} i_n^-e^{\e E_n^-/\hbar\omega}\bigg),
\end{equation}
where the flux dependencies $I(\Phi)$ and $E_n^{\pm}(\Phi)$ have been omitted, and where an energy cutoff ensures that the contribution from deep energy states is exponentially small.
The current per energy level is $i_n^\lambda=-\pd E^\lambda_n/\pd\Phi$,
\begin{equation}\label{ergLevel}
i_n^\pm (\Phi)=
\mp\frac{ I_0 \sin(2\pi\Phi/\Phi_0)}
{\sqrt{\cosh^2(U_0/\hbar\omega)-\cos^2(2\pi\Phi/\Phi_0)}}.
\end{equation}

The PC at half filling is found after carrying the geometric sums in Eq.~(\ref{regCurr}),
\begin{equation}\label{currOdd}
I(\Phi)\big|_{\mu=0}=
\frac{i_0^-}{\pi}\arccos
\bigg[-\frac{\cos(2\pi\Phi/\Phi_0)}{\cosh(U_0/\hbar\omega)}\bigg].
\end{equation}

{\it Effect of electronic filling.} Equivalently, the Fermi energy can be moved at half-integer values $( \mu/\hbar\omega=2N+1/2$), and the PC is computed in a similar fashion.
In this operation, one essentially removes or adds an odd number of fermions in the the system. Therefore it is not surprising that the current-flux characteristic is shifted by half-flux quantum.
For example, to obtain the PC when the Fermi energy moves from the gap at zero energy to the next spectral gap at $\mu/\hbar\omega=1/2$, it is necessary to add the contribution of one more current, $i_0^+$, to the result in Eq.~(\ref{currOdd}).

\begin{figure}[t]
\centering
\includegraphics[width=\columnwidth]{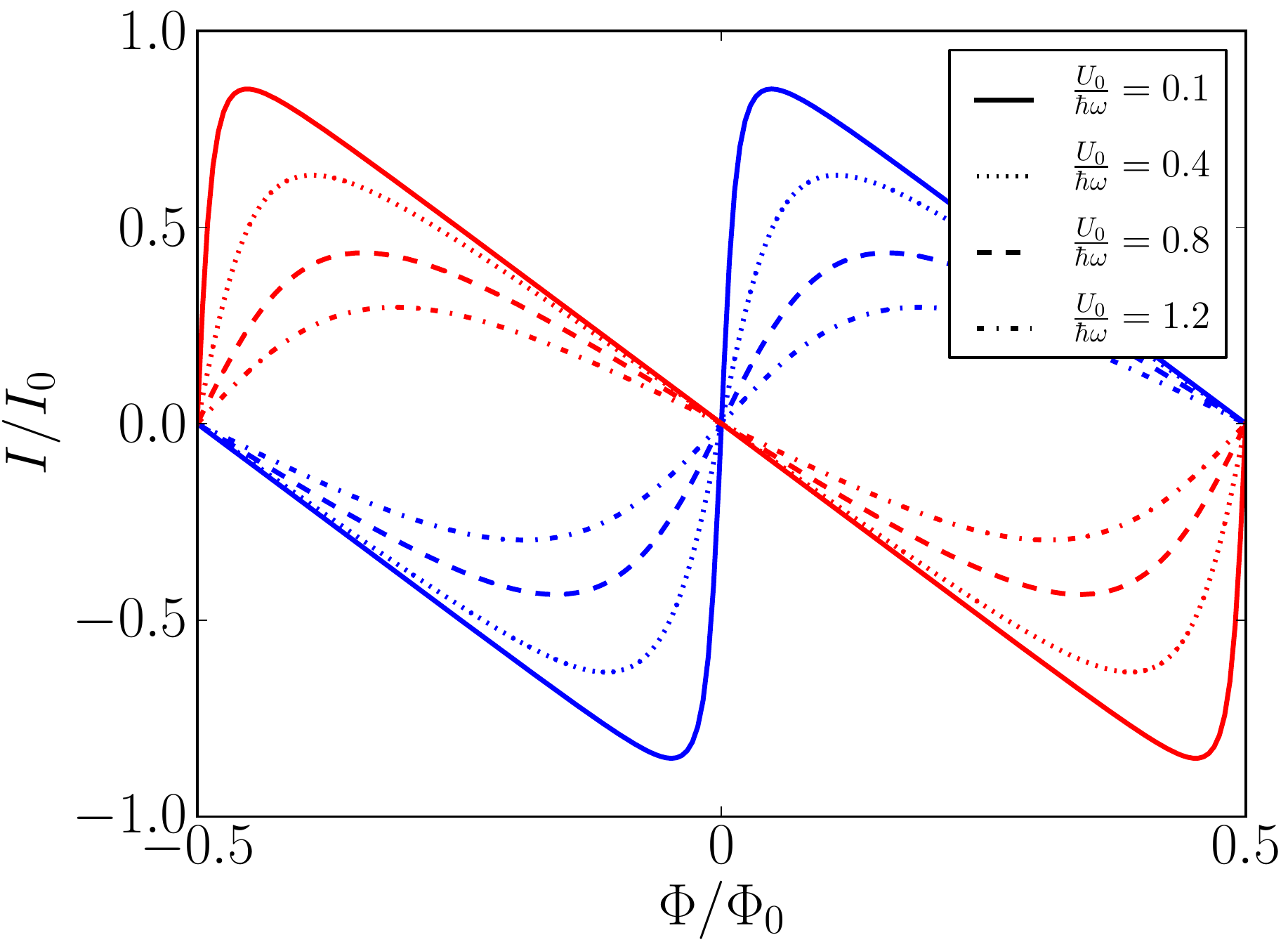}
\caption{(Color online) Regularized persistent current as a function of magnetic flux in the presence of four different strengths $U_0$ for the delta-type magnetic impurity. The blue curves represent cases when the Fermi energy is placed in the gaps at energy $\mu=0$ (modulo $\hbar\omega$), while the red curves, for $\mu=\hbar\omega/2$ (modulo $\hbar\omega$) [see Eq.~(\ref{deltaCurr})]. Parity effects are obtained by varying the Fermi energy.
Increasing the impurity strength leads to exponential suppression in the amplitude for current oscillations as a function of the flux. The unit of current is $I_0=e v_F/L$.}
\label{fig:deltaCurr}
\end{figure}

The general formula for the regularized PC in the presence of a single magnetic scatterer, as a function of the gap which hosts the chemical potential, reads
\begin{eqnarray}\label{deltaCurr}
I(\Phi)&&\bigg|_{\mu=\frac{N\hbar\omega}{2}}=
\frac{I_0 \sin(2\pi\Phi/\Phi_0)}{\sqrt{\cosh^2(U_0/\hbar\omega)-\cos^2(2\pi\Phi/\Phi_0)}}
\notag\\
&&\times\bigg[\frac{(-1)^N}{2}+\frac{1}{\pi}
\arcsin\bigg(\frac{\cos\frac{2\pi\Phi}{\Phi_0}}{\cosh\frac{U_0}{\hbar\omega}}\bigg)\bigg].
\end{eqnarray}
The result clearly indicates that varying the chemical potential by $\hbar\omega/2$ (with $\mu$ in the gaps opened by the magnetic impurity) leads to a shift by $\Phi_0/2$ in the PC-flux characteristic. This ``parity'' effect is represented in Fig.~\ref{fig:deltaCurr} for different impurity strengths. Finally, this result clearly identifies two consequences due to an increasing Dirac-delta magnetic impurity potential: first, a decrease in the amplitude of the current, and second, the destruction of higher harmonics of the signal, which tends to become a simple sine function.

{\it Interpretation.} The decay law of the current as a function of the impurity can be obtained in a heuristic way. As the impurity strength increases, the flux-dependent energy levels become very flat at (half-)integer flux, and the electron velocity decreases. The maximal amplitude of the current will then be given by the point where the velocity is the largest, which is $\Phi/\Phi_0=1/4$. The maximal amplitude of the current does not depend on the parity, so one can fix $N=0$. Then at $\Phi/\Phi_0=1/4$, an expansion of Eq. (\ref{deltaCurr}) in large potential $U_0$ yields the amplitude of the current:
\begin{equation}
I_{\rm max}\sim e^{-L/\xi},\quad\xi=\frac{hv_F}{U_0},
\end{equation}
where we have defined a characteristic length $\xi$ related to the strength $U_0$  of the magnetic impurity. The current amplitude decays exponentially with the the size of the system and the impurity strength. This type of behavior is similar to the one found clean ring with massive Dirac fermions, \cite{Kohno1992} where the typical length was the inverse of the mass of the Dirac fermions. 

\subsubsection{Creutz lattice model}
\label{subsubsec:CrMag}
The addition of a magnetic scatterer modifies the lattice Hamiltonian~(\ref{creutz}) by having an on-site spin coupling at some arbitrary site $J$. The new Hamiltonian, $H_m$, reads
\begin{equation}\label{oneMagCreutz}
H_m=H_{\rm Creutz}+U_0^mc^\dag_J\sigma_1c_J,
\end{equation}
where $U_0^m$ is the strength of magnetic delta potential.

\begin{figure}[t]
\includegraphics[width=\columnwidth]{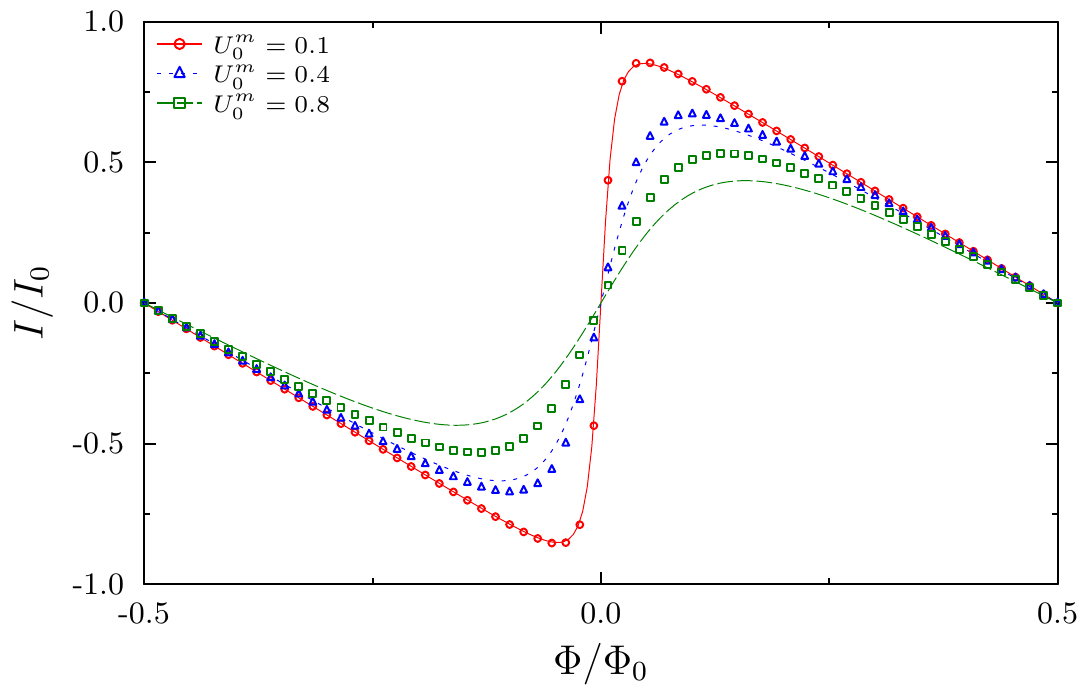}
\caption{(Color online) Current-flux characteristic for three magnetic potentials $U^m_0$ at half filling. The comparison is between the regularized current~(\ref{currOdd}) (represented by lines) for a magnetic Dirac-delta potential in the continuum model of Eq.~(\ref{hamOneMag}), and the numerical results (represented by colored markers) for Creutz model with a single-site potential, Eq.~(\ref{oneMagCreutz}). Following the mapping~(\ref{mapping}), $U_0$ is in units of $\hbar\omega$ in the continuum model and in units of $t$ on the lattice. The analytical results overestimate the impact of the impurity when $U_0^m$ strength becomes comparable to the bandwidth $t$.}
\label{fig:CurrOneComp}
\end{figure}

The numerical results are for an $L=300$ site system. The energies are represented in Fig.~\ref{fig:deltaImpGaps} in units of $\hbar\omega=2\pi ta/L$. When the impurity strength is larger than the  distance between levels, any effects due to the discrete nature of the system are washed out, $U_0^m>2\pi/L$. For large enough systems this condition is always true, and the eigenenergies after scaling with $\hbar\omega$ coincide for different large system sizes.

To compare the analytical results with the numerical simulation, it is necessary to provide the mapping between related quantities. In the analytical case the strength of the impurity is expressed in units of $\hbar\omega=hv_F/L$.
When the system is discretized, the characteristic energy $\hbar\omega$ expressed in terms of lattice parameters reads $2\pi t a/L$, where $a$ is the lattice constant and $t$ the hopping strength. Equivalents for the quantities of interest---Dirac-delta impurity strengths $U_0$ and eigenvalue energies $E$---are obtained similarly:
\begin{equation}\label{mapping}
\hbar\omega\to2\pi t\frac{a}{L},\quad
\frac{U_0}{\hbar\omega}\to\frac{U_0^m}{t},\quad
\frac{E}{\hbar\omega}\to\frac{L}{2\pi a}\frac{E}{t},
\end{equation}
where in the simulation, the hopping strength and lattice constant are $t=1$ and $a=1$.

As mentioned in Sec.~\ref{subsec:scalar}, this simple mapping of the impurity strength between the lattice and continuum model is appropriate in the approximation of infinite bandwidth or for small impurity strength.
This implies that for, e.g., $U_0<\hbar\omega$ (continuum model) and $U_0^m<t$ (lattice model), the numerical results match perfectly the analytical results in Eq.~(\ref{magenergy}). However, at larger values, for $U_0^m$ comparable to $t$, there are deviations from the analytical result.
This approximation remains accurate for small impurity values, and becomes less and less reliable when $U_0$ becomes on the order of $\hbar\omega$ or larger (see Fig.~\ref{fig:deltaImpGaps}).

Similarly one can compare the persistent currents at half filling Eq.~(\ref{currOdd}) obtained by regularizing the infinite Dirac sea pertaining to the helical model in Eq.~(\ref{hamOneMag}) with a magnetic Dirac-delta impurity (see Fig.~\ref{fig:CurrOneComp}). At small disorder potential the match is perfect and, as explained above, becomes less and less reliable when the impurity strength in the lattice model approaches $t$. The analytical results tend to overestimate the impact of a strong single impurity in the system.

\begin{figure}
\includegraphics[width=\columnwidth]{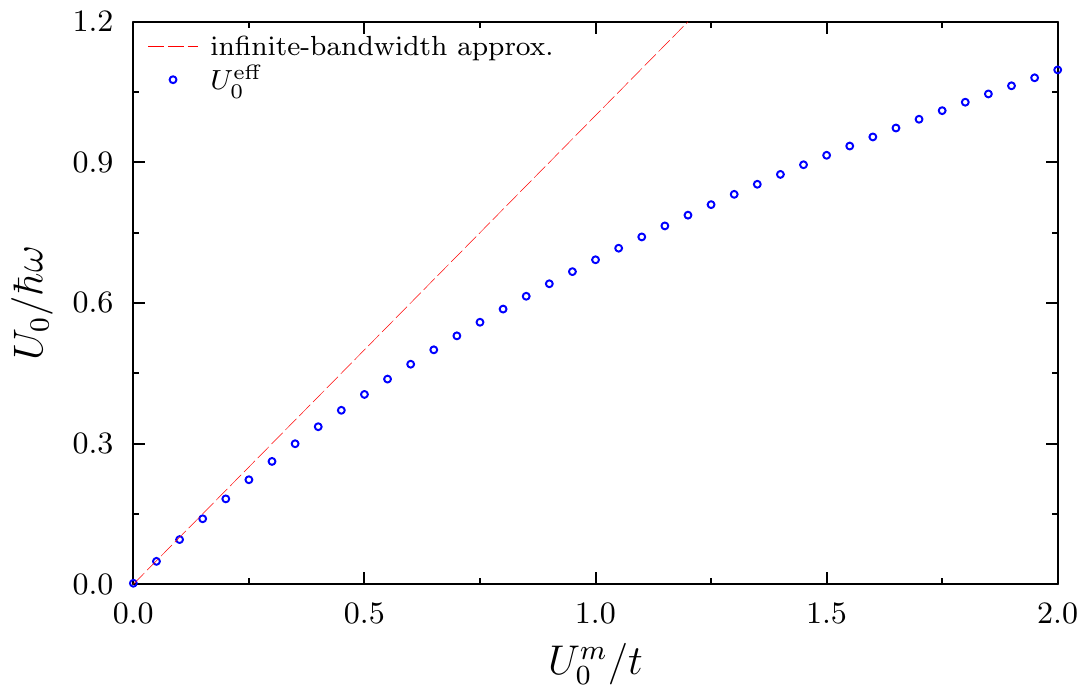}
\caption{(Color online) Impurity strength in the continuum model, $U_0$ (in units of energy-level spacing at the Dirac point), as a function of the strength of the single-site magnetic potential on the lattice, $U_0^m$ (in units of hopping strength $t$). The red dashed line is the mapping suggested in the infinite-bandwidth approximation, which is valid at small impurity strength.
$U_0^{\rm eff}$ is the effective potential in the continuum model, which would reproduce the result on the lattice. The effective potential is described by an universal curve which does not depend on the variation of the $m=g$ parameter.
}
\label{fig:effU}
\end{figure}

For larger impurity strengths, the continuum model impurity strengths needs to be replaced by an effective strength, to recover the lattice results. This is done in the following by considering $U_0$ in the PC expression, Eq.~\ref{deltaCurr}, as a fitting parameter. The result of the fitting procedure is illustrated in Fig.~\ref{fig:effU}.

The effective continuum potential depends on the lattice potential in an universal way: it does not depend on the parameter $m=g$ or on the lattice size $L$. Using the effective potential obtained at zero temperature, it will be possible in Sec.~\ref{sec:temp} to recover a perfect match between the lattice and continuum results even at nonzero temperature.

\subsection{Conclusion}
The results of this section are the following. First, in the case of a single magnetic impurity, both the continuum model and the lattice model are well understood from the fact that backscattering leads to opening of spectral gaps and to a suppression of the persistent current. The analytical formula for the decrease in the persistent current fits well the lattice simulations for small impurity strengths. At large impurity potential, we have numerically obtained the effective potential necessary to match the lattice and continuum models. The renormalized continuum potential does not depend on lattice or parameter  $m=g$ variation (or on temperature; see Sec.~\ref{sec:temp}).

Second, the consideration of a scalar impurity has revealed that the persistent current in the lattice model equally leads to a decrease in the persistent current. The lack of TRS in the lattice model allows opening spectral gaps at (half-)integer flux. Combined with the symmetry constraints of the PTRS, this leads to zero current at these values of the flux. Then the impurity has smoothed the current discontinuities, and the PC will exponentially decay with the impurity strength.

\section{Temperature effects}
\label{sec:temp}
Finite temperature suppresses the phase coherence of the electronic wave functions.
This in turn implies that quantum interference effects such as the Aharonov-Bohm phase and the persistent currents are suppressed when temperature is increased. In the context of quantum rings with nonrelativistic fermions, the decay of persistent currents under temperature was first studied in  Refs.~\onlinecite{Buettiker1985,Cheung1988}. Moreover, PC fluctuations can survive at (relative) higher temperatures, even for vanishing average PC.\cite{Moskalets2010} 

The present section investigates the temperature dependence of the average persistent current within the two Dirac models studied in this paper.
It considers first the clean-ring models (Sec.~\ref{sub:cleanfiniteT}), and afterward rings with a single magnetic impurity (Sec.~\ref{sub:impurityfiniteT}). In both cases, the system is treated in the grand-canonical ensemble, at fixed chemical potential $\mu$ and temperature $T$.

For the continuum helical model, the current-flux relation $I(\Phi)$ is derived analytically using Eq.~(\ref{currentDEF}) and the ultraviolet regularization introduced in Sec.~\ref{subsec:magnetic}. Numerical lattice simulations on the Creutz model agree with the analytical results for the helical Dirac model.

\subsection{Clean ring.}
\label{sub:cleanfiniteT}
At finite temperature, there is a single relevant energy scale in the clean ring: the energy-level spacing at the Dirac point, $\hbar\omega=2\pi\hbar v_F/L$.
This determines a characteristic temperature $T^*$ for metals, which is proportional to the level spacing,~\cite{Cheung1988}
\begin{equation}\label{charTemp}
T^*=\frac{\hbar v_F}{\pi k_B L},
\end{equation}
where $k_B$ is the Boltzmann constant. 

{\it Continuum helical model.} Let us first consider the helical model defined by Eq.~(\ref{hamiltonianconti}).
Using the general formula Eq.~(\ref{currentSUMfinite}) with the clean ring spectrum Eq.~(\ref{spectrumconti}), one obtains the persistent current
\begin{equation}\label{currTemp}
I^\pm=\lim_{\e\to 0} \sum_{\sigma =\pm} \,  \sum_{n=-\infty}^\infty
f(E^\sigma_n)
i_n^\sigma e^{\e\frac{E_n^{\sigma} - \mu}{\hbar\omega}}
%e^{\beta (E^\sigma_n-\mu)}+1
,
\end{equation}
where $\e$ is a small positive constant used to fix an ultraviolet energy cutoff. All states below the chemical potential are suppressed. The function $f(x)$ is the Fermi-Dirac distribution. 
The currents carried by each level are $i^\sigma_n=-\pd E^\sigma_n/\pd\Phi$.

The Fourier expansion of the regularized total PC reads (see the Appendix)
\begin{equation}\label{currTempRes1}
I(\Phi)=\sum_{m=1}^\infty \mc I_m(T,\mu) \sin\big(2\pi m\frac{\Phi}{\Phi_0}\big)
\end{equation}
where the coefficients are given by
\begin{equation}\label{currTempRes2}
\mc I_m(T,\mu)=\frac{2I_0T}{\pi T^*}
\frac{\cos\big(2\pi m\frac{\mu}{\hbar\omega})}{\sinh(mT/T^*)}.
\end{equation}
Above the characteristic temperature $T^*$, the Fourier components $\mc I_m$ decay exponentially, and the PC-flux characteristic is very well approximated by the first harmonics in the Fourier expansion. In contrast, close to zero temperature, there are PC discontinuities ($2I_0$ jumps for $\mu=0$ or $\hbar\omega/2$ (modulo $\hbar\omega$)) which are approximated only by summing many Fourier components. Note that in the zero-temperature limit, and at zero chemical potential, the result in Eq.~(\ref{currClean}) is recovered, $\mc I_m(T=0,\mu=0)=2I_0/(\pi m)$.

The cosine dependence on the chemical potential encompasses the parity effects that were studied in the previous sections. A change by $\hbar\omega/2$ in the chemical potential $\mu$ is equivalent to a shift of the flux $\Phi$ by half-flux quantum.

The PC given by Eqs.~(\ref{currTempRes1}) and (\ref{currTempRes2}) is identical to the one obtained in Ref.~\onlinecite{Cheung1988} for fermions with quadratic dispersion in metallic rings. 

\begin{figure}[t]
\includegraphics[width=\columnwidth]{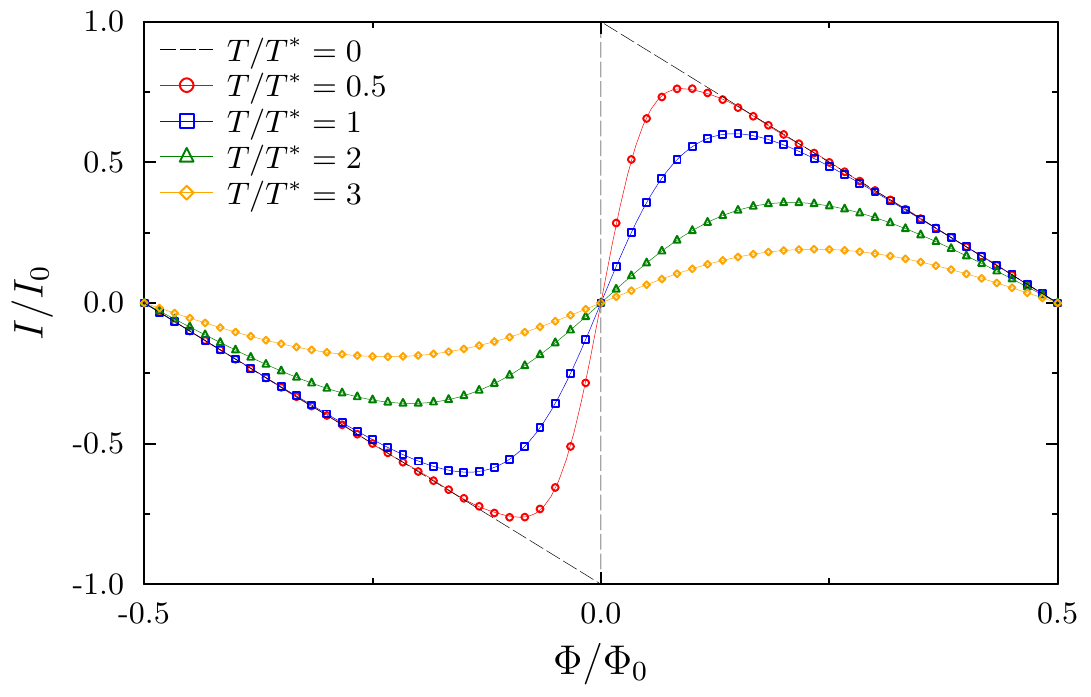}
\caption{(Color online) Temperature effects in the lattice (Creutz) model and the continuum model at zero chemical potential. The straight line represents analytical expression for the current, summing the first ten Fourier components of the current given by Eqs.~(\ref{currTempRes1}) and~(\ref{currTempRes2}), and the markers represent numerical simulations in the lattice (Creutz) model. The ring circumference is $L=300 a$.}
\label{fig:temp}
\end{figure}

{\it Creutz lattice model.} In the lattice model, the PC at finite temperature is also obtained by using the general formula Eq.~(\ref{currentSUMfinite}) with the spectrum obtained by diagonalization of the lattice Hamiltonian Eq. (\ref{creutz}). The characteristic temperature $T^*$ is given in Eq.~(\ref{charTemp}), with the Fermi velocity $\hbar v_F=ta$. Figure~\ref{fig:temp} shows that the the current of the analytical model fits perfectly the Creutz lattice model for a range of temperatures above and below $T^*$.

\subsection{Single magnetic impurity.}
\label{sub:impurityfiniteT}
We now turn to the ring in presence of a single magnetic impurity, which mixes spins, and open gaps at all the crossings in the energy-flux spectrum. At finite temperature, the PC is given by the general formula Eq.~(\ref{currentSUMfinite}) used with the spectrum Eqs.~(\ref{magenergy}) and (\ref{effectivephase}) of the ring in presence of a magnetic impurity:
\begin{equation}
I=\lim_{\e\to0}\sum_{\lambda=\pm}\sum_{n=-\infty}^\infty
f(E_n^\lambda)
i_n^\lambda e^{\e(E_n^\lambda-\mu)/\hbar\omega},
\end{equation}
where $\lambda=\pm$ are the level (not spin) indices introduced in Sec.~\ref{subsec:magnetic}. The energy eigenstates $E_n^\lambda(\Phi)$ are given in Eqs.~(\ref{magenergy}) and~(\ref{effectivephase}), and the level currents $i_n^\lambda(\Phi)$, in Eq.~(\ref{ergLevel}).

Following the regularization procedure detailed in the Appendix, it follows readily that the PC is
\begin{equation}\label{impCurrTemp}
I=i_0^-\sum_{m=1}^\infty \frac{2T}{\pi T^*}
\frac{\cos(2\pi m\frac{\mu}{\hbar\omega})}{\sinh(mT/T^*)}
\sin(2\pi m\varphi).
\end{equation}
In comparison with the PC in the clean ring case from Eq.~(\ref{currTempRes1},\ref{currTempRes2}), the flux $\Phi/\Phi_0$ in the Fourier harmonics is naturally replaced by the effective phase, $\varphi(\Phi)$, determined from the scattering of electrons on the magnetic potential (see Sec.~\ref{subsec:magnetic}). Also note that instead of the current carried by an eigenstate in the clean ring near the Dirac point, $I_0$, the formula contains the current $i_0^-$ in one of the energy-flux bands which were formed by the impurity potential. These two quantities depend on the strength of the magnetic Dirac-delta impurity, $U_0/\hbar\omega$. The expression for $i_0^-$ and $\varphi(\Phi)$ are recalled here for the reader's convenience: 
\begin{equation}
i_0^- (\Phi)=
\frac{ I_0 \sin(2\pi \Phi/\Phi_0)}
{\sqrt{\cosh^2(U_0/\hbar\omega)-\cos^2(2\pi\Phi/\Phi_0)}},
\end{equation}
and
\begin{equation}
\varphi=\frac{1}{2\pi}\arccos\bigg[
\frac{\cos(2\pi\Phi/\Phi_0)}{\cosh(U_0/\hbar\omega)}
\bigg].
\end{equation}

\begin{figure}[t]
\includegraphics[width=\columnwidth]{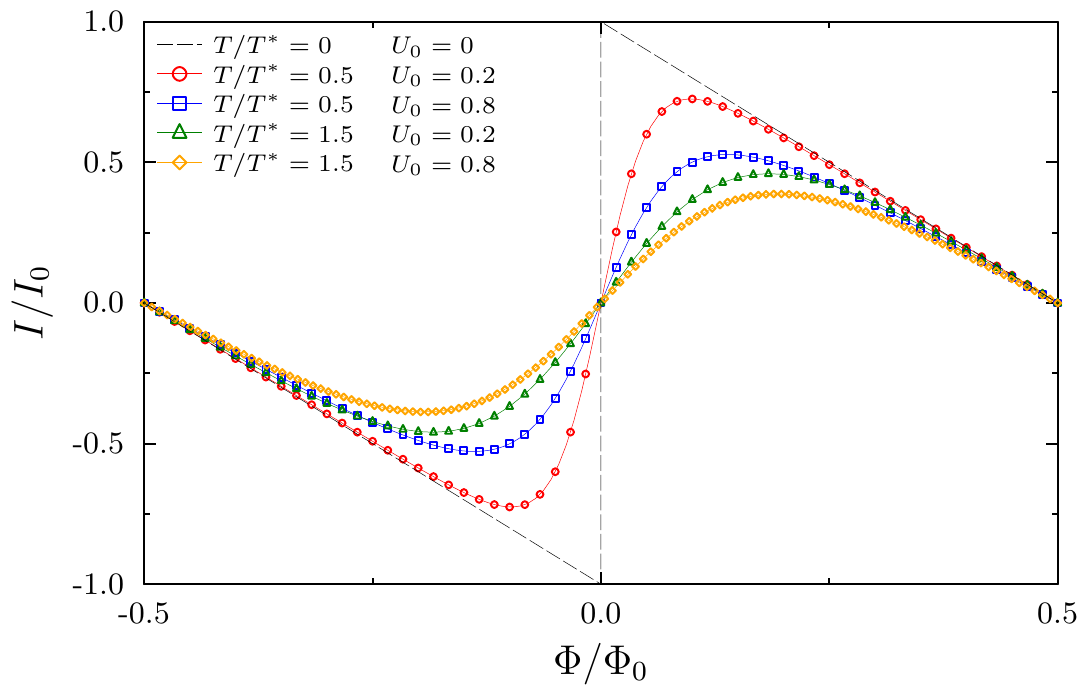}
\caption{(Color online) Decay of the persistent current under the combined influence of temperature and magnetic potential $U_0$. The magnetic impurity strength in the lattice model is measured in units of $t$ on the lattice and listed in the legend. In the continuum model, the effective potential is given by the curve in Fig.~\ref{fig:effU} (in units of $\hbar\omega$). Straight lines represent the analytical prediction from the helical continuum model~(\ref{hamiltonianconti}), and the markers, the exact-diagonalization results from Creutz model~(\ref{creutz}). 
The system size is $L=300 a$.
}
\label{fig:tempMag}
\end{figure}

In the limit of zero temperature, and chemical potential $\mu\in\{0,\frac{\hbar\omega}{2}\}$ $(\text{mod } \hbar\omega)$, one recovers the previous result for the PC in Eq.~(\ref{deltaCurr}).

The lattice simulations match remarkably the analytical results for small impurity strength $U_0$, regardless of temperature. Any divergence of numerics from the result in Eq.~(\ref{impCurrTemp}) is due to the infinite-bandwidth approximation, and not to temperature effects. As in the case of spinless chiral fermions,\cite{Gogolin2004} the analytical result in the infinite-bandwidth approximation overestimates the effect of the impurity. 
To cure this problem, instead of the simple mapping between the impurity strength in the continuum model and lattice model, $U_0/\hbar\omega\to U_0/t$, it is better to use an effective potential in the continuum. This renormalized potential was determined before in the zero-temperature case (Fig.~\ref{fig:effU}). The analytical formula~in Eq.~(\ref{impCurrTemp}), supplemented by the effective potential, can account perfectly for the lattice result.
Figure~\ref{fig:tempMag} represents this match between lattice and analytical results at small and large impurity potential, for temperatures above and below $T^*$. This shows how both temperature and the magnetic potential conspire to decrease the PC. But most importantly, it shows that fitting a single quantity, the continuum $U_0$, is enough to map the lattice and continuum results, independent of system parameters (arbitrary $m=g$), system size, or temperature. This proves the utility of the analytical expression for persistent currents (Eq.~\ref{impCurrTemp}) in the context of Dirac fermion rings with a single magnetic impurity.

\section{Conclusion}
\label{sec:conclusion}
We have considered the persistent currents in two different models of strictly one-dimensional Dirac fermions. The two models share the same low-energy dispersion, but they have very different spectra at high energy and  different wave functions, even at low energy. The first model, the Dirac helical model, is defined on a continuous ring and it allows us to compute persistent currents analytically, using an ultraviolet regularization of the unbound negative Dirac sea. The second model, the Creutz model, is defined on a lattice and allows well-controlled numerical calculations of the flux-dependent spectrum, total energy, and persistent currents.

In the case of a clean ring, the two models lead to the same persistent current, $I(\Phi)$, which is a piecewise linear function of magnetic flux $\Phi$ (at zero temperature) with period $\Phi_0=h/e$ (inset of Fig.~\ref{fig:effenergy}). This type of current-flux relation is also exactly the one obtained for nonrelativistic fermions described by a simple parabolic dispersion relation in one-dimensional rings.\cite{Cheung1988,Cheung1989} This shows a remarkable independence of the persistent current on the details of the band dispersion. However, on the lattice system it was also possible to tune between one and two massless Dirac fermions, which results in a doubling (or halving) of the total amplitude of the current.

The main part of the paper is devoted to the effect of a single impurity on the persistent current, $I(\Phi)$, flowing in such Dirac rings using both continuum and lattice models. The cases of nonmagnetic and magnetic impurities are contrasted.

First, the effect of a scalar spin-independent impurity is treated analytically in the continuum helical model. Then adding a single nonmagnetic impurity cannot create backscattering, and no effects on the persistent current are predicted within the ideal helical model. However, the lattice model is more sensitive to such nonmagnetic impurity because time-reversal symmetry is broken in the Creutz model. As a consequence, gaps can occur in the energy-flux spectrum, thereby leading to a decrease of the persistent current. 

Second, it is shown that a magnetic impurity is more harmful to the persistent current than the nonmagnetic impurity in both continuum and lattice models. The PC is computed analytically for a single Dirac-delta magnetic impurity in the helical Dirac model, both at zero and finite temperature. The decay of the current due to impurity or temperature effects was compared to the lattice simulations. The analytical formulas agree with lattice simulations in the limit of a small impurity strength compared to the bandwidth. For large impurity strength, a renormalized continuum potential is necessary to ensure the match between the two models.

In perspective, it would be of great value to extend the present study to different geometries including cylinders, nanowires, and disks of topological insulators hosting various types of helical edge states. Moreover, extension to cases with multiple magnetic and nonmagnetic scatterers is wanting.
This direct extension of the present study would uncover the localization physics of Wilson-Dirac fermions.\cite{Wimmer2010} Due to TRS breaking in the lattice model, one expects that even an on-site random scalar potential would produce an exponential decay of the persistent current.

\medskip

We thank I.~Herbut, K.-I.~Imura, G.~Montambaux, and P.~Ribeiro for enlightening discussions. J.C.\ acknowledges support from EU/FP7 under contract TEMSSOC. This work was supported by the French ANR through Project No.\
2010-BLANC-041902 (ISOTOP). This research has been  supported by the Hungarian Scientific
Research Funds No.~K101244, No.~K105149, and No.~CNK80991, by the Bolyai Program of the Hungarian Academy of Sciences, and partially by ERC Grant No.~ERC-259374-Sylo.

\appendix*
\section{Persistent current in a clean Dirac fermion ring at finite temperature}
Let us consider the helical Dirac model represented by the Hamiltonian in Eq.~(\ref{hamiltonianconti}).
The persistent current at finite temperature is a sum over the currents carried by each eigenstate, weighted by the Fermi-Dirac distribution. 
The system contains an infinite number of occupied states. To extract meaningful physical results it is necessary to apply an ultraviolet regularization of the currents.

Then, at arbitrary chemical potential $\mu$, the currents for spin-up states $(+)$ and spin-down states $(-)$ read
\begin{equation}
I^\pm=\sum_{n=-\infty}^\infty
\frac{i_n^\pm\exp[\e(E_n^{\pm}-\mu)/\hbar\omega]}{e^{\beta(E^\pm_n-\mu)}+1},
\end{equation}
where the constant $\e$ is an infinitesimal positive number. The constant $\e$ ensures an energy cutoff for deep states below the chemical potential $\mu$.
The currents carried by each level are $i_n^\pm=-\pd E^\pm_n/\pd\Phi$, with the energy eigenstates $E_n^\pm=\pm\hbar\omega(n+\Phi/\Phi_0)$.

The Poisson summation formula can be used to further simplify the problem,
\begin{equation}
\sum_{n=-\infty}^\infty f(n)
=\sum_{m=-\infty}^\infty\int_{-\infty}^\infty dx f(x)e^{2\pi i m x}.
\end{equation}

Let us introduce the variable $y=\hbar\omega\beta(x+\Phi/\Phi_0)-\beta\mu$ in the $I^+$ expression, and $y=\hbar\omega\beta(-x-\Phi/\Phi_0)-\beta\mu$ in the $I^-$ expression. Then the currents read
\begin{equation}
\frac{I^\pm}{I_0}=\mp\sum_{m=-\infty}^\infty \int_{-\infty}^\infty dy 
\frac{\exp\big[\frac{(\e\pm 2\pi im) y}{\hbar\omega\beta}\big]}{e^{y}+1} \frac{e^{-2\pi i m(\frac{\Phi}{\Phi_0}\mp\frac{\mu}{\hbar\omega})}}{\hbar\omega\beta},
\end{equation}
where the current is expressed in units of $I_0=\hbar\omega/\Phi_0=ev_F/L$.
The presence of an infinitesimal positive $\e$ ensures the convergence of the integral over $y$. The integrals can be carried out exactly by contour integration, yielding
\begin{equation}
\frac{I^\pm}{I_0}=\mp\sum_{m=-\infty}^\infty
\frac{T}{2\pi T^*}\csc
\big[\frac{T}{T^*}\big(\frac{\e}{2\pi}\pm i m
\big)\big]
e^{-2\pi i m(\frac{\Phi}{\Phi_0}\mp\frac{\mu}{\hbar\omega})},
\end{equation}
where for convenience the following notation was introduced:
\begin{equation}
T^*=\frac{\hbar\omega}{2\pi^2 k_B}=\frac{\hbar v_F}{\pi k_B L}.
\end{equation}
The temperature $T^*$ is a characteristic temperature for system and it is determined by the energy level spacing at the Dirac point in zero flux, $\hbar\omega$.

The total current is defined as $I=I^++I^-$. Expanding in $\e$ allows one to single out the contribution independent of the cutoff,
\begin{equation}
\frac{I}{I_0}=\sum_{m=1}^\infty\frac{2T}{\pi T^*}
\frac{\cos\big(2\pi m\frac{\mu}{\hbar\omega})}{\sinh(mT/T^*)}\sin(2\pi m\frac{\Phi}{\Phi_0})+O(\e)
\end{equation}
The cutoff can now be safely taken to zero to give the physical result. The above formula is nothing but the Fourier series of an odd quantity in the flux $\Phi$. The PC's Fourier components $\mc I_m$, in units of the maximal current carried by an eigenstate $I_0$, read
\begin{equation}
\frac{\mc I_m}{I_0}=\frac{2T}{\pi T^*}
\frac{\cos\big(2\pi m\frac{\mu}{\hbar\omega})}{\sinh(mT/T^*)}.
\end{equation}
The PC in the Dirac helical model~(\ref{hamiltonianconti}) proves to be identical to the PC expression for electrons with quadratic dispersion in small metallic rings.\cite{Cheung1988}

\bibliographystyle{apsrev4-1}
\bibliography{bibl}

\begin{thebibliography}{56}%
\makeatletter
\providecommand \@ifxundefined [1]{%
 \@ifx{#1\undefined}
}%
\providecommand \@ifnum [1]{%
 \ifnum #1\expandafter \@firstoftwo
 \else \expandafter \@secondoftwo
 \fi
}%
\providecommand \@ifx [1]{%
 \ifx #1\expandafter \@firstoftwo
 \else \expandafter \@secondoftwo
 \fi
}%
\providecommand \natexlab [1]{#1}%
\providecommand \enquote  [1]{``#1''}%
\providecommand \bibnamefont  [1]{#1}%
\providecommand \bibfnamefont [1]{#1}%
\providecommand \citenamefont [1]{#1}%
\providecommand \href@noop [0]{\@secondoftwo}%
\providecommand \href [0]{\begingroup \@sanitize@url \@href}%
\providecommand \@href[1]{\@@startlink{#1}\@@href}%
\providecommand \@@href[1]{\endgroup#1\@@endlink}%
\providecommand \@sanitize@url [0]{\catcode `\\12\catcode `\$12\catcode
  `\&12\catcode `\#12\catcode `\^12\catcode `\_12\catcode `\%12\relax}%
\providecommand \@@startlink[1]{}%
\providecommand \@@endlink[0]{}%
\providecommand \url  [0]{\begingroup\@sanitize@url \@url }%
\providecommand \@url [1]{\endgroup\@href {#1}{\urlprefix }}%
\providecommand \urlprefix  [0]{URL }%
\providecommand \Eprint [0]{\href }%
\providecommand \doibase [0]{http://dx.doi.org/}%
\providecommand \selectlanguage [0]{\@gobble}%
\providecommand \bibinfo  [0]{\@secondoftwo}%
\providecommand \bibfield  [0]{\@secondoftwo}%
\providecommand \translation [1]{[#1]}%
\providecommand \BibitemOpen [0]{}%
\providecommand \bibitemStop [0]{}%
\providecommand \bibitemNoStop [0]{.\EOS\space}%
\providecommand \EOS [0]{\spacefactor3000\relax}%
\providecommand \BibitemShut  [1]{\csname bibitem#1\endcsname}%
\let\auto@bib@innerbib\@empty
%</preamble>
\bibitem [{\citenamefont {Imry}(2008)}]{Imry2008}%
  \BibitemOpen
  \bibfield  {author} {\bibinfo {author} {\bibfnamefont {Y.}~\bibnamefont
  {Imry}},\ }\href@noop {} {\emph {\bibinfo {title} {Introduction to Mesoscopic
  Physics}}},\ \bibinfo {edition} {2nd}\ ed.\ (\bibinfo  {publisher} {Oxford
  University Press, New York},\ \bibinfo {year} {2008})\BibitemShut {NoStop}%
\bibitem [{\citenamefont {Akkermans}\ and\ \citenamefont
  {Montambaux}(2011)}]{AkkM2011}%
  \BibitemOpen
  \bibfield  {author} {\bibinfo {author} {\bibfnamefont {E.}~\bibnamefont
  {Akkermans}}\ and\ \bibinfo {author} {\bibfnamefont {G.}~\bibnamefont
  {Montambaux}},\ }\href {http://amazon.com/o/ASIN/0521349478/} {\emph
  {\bibinfo {title} {Mesoscopic Physics of Electrons and Photons}}},\ \bibinfo
  {edition} {reissue}\ ed.\ (\bibinfo  {publisher} {Cambridge University
  Press},\ \bibinfo {year} {2011})\BibitemShut {NoStop}%
\bibitem [{\citenamefont {Byers}\ and\ \citenamefont {Yang}(1961)}]{Byers1961}%
  \BibitemOpen
  \bibfield  {author} {\bibinfo {author} {\bibfnamefont {N.}~\bibnamefont
  {Byers}}\ and\ \bibinfo {author} {\bibfnamefont {C.~N.}\ \bibnamefont
  {Yang}},\ }\href {\doibase 10.1103/PhysRevLett.7.46} {\bibfield  {journal}
  {\bibinfo  {journal} {Phys. Rev. Lett.}\ }\textbf {\bibinfo {volume} {7}},\
  \bibinfo {pages} {46} (\bibinfo {year} {1961})}\BibitemShut {NoStop}%
\bibitem [{\citenamefont {B\"uttiker}\ \emph {et~al.}(1983)\citenamefont
  {B\"uttiker}, \citenamefont {Imry},\ and\ \citenamefont
  {Landauer}}]{Buettiker1983}%
  \BibitemOpen
  \bibfield  {author} {\bibinfo {author} {\bibfnamefont {M.}~\bibnamefont
  {B\"uttiker}}, \bibinfo {author} {\bibfnamefont {Y.}~\bibnamefont {Imry}}, \
  and\ \bibinfo {author} {\bibfnamefont {R.}~\bibnamefont {Landauer}},\ }\href
  {\doibase 10.1016/0375-9601(83)90011-7} {\bibfield  {journal} {\bibinfo
  {journal} {Phys. Lett. A}\ }\textbf {\bibinfo {volume} {96}},\ \bibinfo
  {pages} {365 } (\bibinfo {year} {1983})}\BibitemShut {NoStop}%
\bibitem [{\citenamefont {Ambegaokar}\ and\ \citenamefont
  {Eckern}(1990)}]{Ambegaokar1990}%
  \BibitemOpen
  \bibfield  {author} {\bibinfo {author} {\bibfnamefont {V.}~\bibnamefont
  {Ambegaokar}}\ and\ \bibinfo {author} {\bibfnamefont {U.}~\bibnamefont
  {Eckern}},\ }\href {\doibase 10.1103/PhysRevLett.65.381} {\bibfield
  {journal} {\bibinfo  {journal} {Phys. Rev. Lett.}\ }\textbf {\bibinfo
  {volume} {65}},\ \bibinfo {pages} {381} (\bibinfo {year} {1990})}\BibitemShut
  {NoStop}%
\bibitem [{\citenamefont {Abraham}\ and\ \citenamefont
  {Berkovits}(1993)}]{Berkovits:1993}%
  \BibitemOpen
  \bibfield  {author} {\bibinfo {author} {\bibfnamefont {M.}~\bibnamefont
  {Abraham}}\ and\ \bibinfo {author} {\bibfnamefont {R.}~\bibnamefont
  {Berkovits}},\ }\href {\doibase 10.1103/PhysRevLett.70.1509} {\bibfield
  {journal} {\bibinfo  {journal} {Phys. Rev. Lett.}\ }\textbf {\bibinfo
  {volume} {70}},\ \bibinfo {pages} {1509} (\bibinfo {year}
  {1993})}\BibitemShut {NoStop}%
\bibitem [{\citenamefont {M\"uller-Groeling}\ and\ \citenamefont
  {Weidenm\"uller}(1994)}]{Muller:1994}%
  \BibitemOpen
  \bibfield  {author} {\bibinfo {author} {\bibfnamefont {A.}~\bibnamefont
  {M\"uller-Groeling}}\ and\ \bibinfo {author} {\bibfnamefont {H.~A.}\
  \bibnamefont {Weidenm\"uller}},\ }\href {\doibase 10.1103/PhysRevB.49.4752}
  {\bibfield  {journal} {\bibinfo  {journal} {Phys. Rev. B}\ }\textbf {\bibinfo
  {volume} {49}},\ \bibinfo {pages} {4752} (\bibinfo {year}
  {1994})}\BibitemShut {NoStop}%
\bibitem [{\citenamefont {Bary-Soroker}\ \emph {et~al.}(2008)\citenamefont
  {Bary-Soroker}, \citenamefont {Entin-Wohlman},\ and\ \citenamefont
  {Imry}}]{Bary-Soroker2008}%
  \BibitemOpen
  \bibfield  {author} {\bibinfo {author} {\bibfnamefont {H.}~\bibnamefont
  {Bary-Soroker}}, \bibinfo {author} {\bibfnamefont {O.}~\bibnamefont
  {Entin-Wohlman}}, \ and\ \bibinfo {author} {\bibfnamefont {Y.}~\bibnamefont
  {Imry}},\ }\href {\doibase 10.1103/PhysRevLett.101.057001} {\bibfield
  {journal} {\bibinfo  {journal} {Phys. Rev. Lett.}\ }\textbf {\bibinfo
  {volume} {101}},\ \bibinfo {pages} {057001} (\bibinfo {year}
  {2008})}\BibitemShut {NoStop}%
\bibitem [{\citenamefont {Cheung}\ \emph {et~al.}(1988)\citenamefont {Cheung},
  \citenamefont {Gefen}, \citenamefont {Riedel},\ and\ \citenamefont
  {Shih}}]{Cheung1988}%
  \BibitemOpen
  \bibfield  {author} {\bibinfo {author} {\bibfnamefont {H.-F.}\ \bibnamefont
  {Cheung}}, \bibinfo {author} {\bibfnamefont {Y.}~\bibnamefont {Gefen}},
  \bibinfo {author} {\bibfnamefont {E.~K.}\ \bibnamefont {Riedel}}, \ and\
  \bibinfo {author} {\bibfnamefont {W.-H.}\ \bibnamefont {Shih}},\ }\href
  {\doibase 10.1103/PhysRevB.37.6050} {\bibfield  {journal} {\bibinfo
  {journal} {Phys. Rev. B}\ }\textbf {\bibinfo {volume} {37}},\ \bibinfo
  {pages} {6050} (\bibinfo {year} {1988})}\BibitemShut {NoStop}%
\bibitem [{\citenamefont {Cheung}\ \emph {et~al.}(1989)\citenamefont {Cheung},
  \citenamefont {Riedel},\ and\ \citenamefont {Gefen}}]{Cheung1989}%
  \BibitemOpen
  \bibfield  {author} {\bibinfo {author} {\bibfnamefont {H.-F.}\ \bibnamefont
  {Cheung}}, \bibinfo {author} {\bibfnamefont {E.~K.}\ \bibnamefont {Riedel}},
  \ and\ \bibinfo {author} {\bibfnamefont {Y.}~\bibnamefont {Gefen}},\ }\href
  {\doibase 10.1103/PhysRevLett.62.587} {\bibfield  {journal} {\bibinfo
  {journal} {Phys. Rev. Lett.}\ }\textbf {\bibinfo {volume} {62}},\ \bibinfo
  {pages} {587} (\bibinfo {year} {1989})}\BibitemShut {NoStop}%
\bibitem [{\citenamefont {Viefers}\ \emph {et~al.}(2004)\citenamefont
  {Viefers}, \citenamefont {Koskinen}, \citenamefont {Deo},\ and\ \citenamefont
  {Manninen}}]{Viefers2004}%
  \BibitemOpen
  \bibfield  {author} {\bibinfo {author} {\bibfnamefont {S.}~\bibnamefont
  {Viefers}}, \bibinfo {author} {\bibfnamefont {P.}~\bibnamefont {Koskinen}},
  \bibinfo {author} {\bibfnamefont {P.~S.}\ \bibnamefont {Deo}}, \ and\
  \bibinfo {author} {\bibfnamefont {M.}~\bibnamefont {Manninen}},\ }\href
  {\doibase 10.1016/j.physe.2003.08.076} {\bibfield  {journal} {\bibinfo
  {journal} {Physica E}\ }\textbf {\bibinfo {volume} {21}},\ \bibinfo {pages}
  {1 } (\bibinfo {year} {2004})}\BibitemShut {NoStop}%
\bibitem [{\citenamefont {L\'evy}\ \emph {et~al.}(1990)\citenamefont {L\'evy},
  \citenamefont {Dolan}, \citenamefont {Dunsmuir},\ and\ \citenamefont
  {Bouchiat}}]{Levy1990}%
  \BibitemOpen
  \bibfield  {author} {\bibinfo {author} {\bibfnamefont {L.~P.}\ \bibnamefont
  {L\'evy}}, \bibinfo {author} {\bibfnamefont {G.}~\bibnamefont {Dolan}},
  \bibinfo {author} {\bibfnamefont {J.}~\bibnamefont {Dunsmuir}}, \ and\
  \bibinfo {author} {\bibfnamefont {H.}~\bibnamefont {Bouchiat}},\ }\href
  {\doibase 10.1103/PhysRevLett.64.2074} {\bibfield  {journal} {\bibinfo
  {journal} {Phys. Rev. Lett.}\ }\textbf {\bibinfo {volume} {64}},\ \bibinfo
  {pages} {2074} (\bibinfo {year} {1990})}\BibitemShut {NoStop}%
\bibitem [{\citenamefont {Chandrasekhar}\ \emph {et~al.}(1991)\citenamefont
  {Chandrasekhar}, \citenamefont {Webb}, \citenamefont {Brady}, \citenamefont
  {Ketchen}, \citenamefont {Gallagher},\ and\ \citenamefont
  {Kleinsasser}}]{Chandrasekhar1991}%
  \BibitemOpen
  \bibfield  {author} {\bibinfo {author} {\bibfnamefont {V.}~\bibnamefont
  {Chandrasekhar}}, \bibinfo {author} {\bibfnamefont {R.~A.}\ \bibnamefont
  {Webb}}, \bibinfo {author} {\bibfnamefont {M.~J.}\ \bibnamefont {Brady}},
  \bibinfo {author} {\bibfnamefont {M.~B.}\ \bibnamefont {Ketchen}}, \bibinfo
  {author} {\bibfnamefont {W.~J.}\ \bibnamefont {Gallagher}}, \ and\ \bibinfo
  {author} {\bibfnamefont {A.}~\bibnamefont {Kleinsasser}},\ }\href {\doibase
  10.1103/PhysRevLett.67.3578} {\bibfield  {journal} {\bibinfo  {journal}
  {Phys. Rev. Lett.}\ }\textbf {\bibinfo {volume} {67}},\ \bibinfo {pages}
  {3578} (\bibinfo {year} {1991})}\BibitemShut {NoStop}%
\bibitem [{\citenamefont {Mailly}\ \emph {et~al.}(1993)\citenamefont {Mailly},
  \citenamefont {Chapelier},\ and\ \citenamefont {Benoit}}]{Mailly1993}%
  \BibitemOpen
  \bibfield  {author} {\bibinfo {author} {\bibfnamefont {D.}~\bibnamefont
  {Mailly}}, \bibinfo {author} {\bibfnamefont {C.}~\bibnamefont {Chapelier}}, \
  and\ \bibinfo {author} {\bibfnamefont {A.}~\bibnamefont {Benoit}},\ }\href
  {\doibase 10.1103/PhysRevLett.70.2020} {\bibfield  {journal} {\bibinfo
  {journal} {Phys. Rev. Lett.}\ }\textbf {\bibinfo {volume} {70}},\ \bibinfo
  {pages} {2020} (\bibinfo {year} {1993})}\BibitemShut {NoStop}%
\bibitem [{\citenamefont {Bleszynski-Jayich}\ \emph {et~al.}(2009)\citenamefont
  {Bleszynski-Jayich}, \citenamefont {Shanks}, \citenamefont {Peaudecerf},
  \citenamefont {Ginossar}, \citenamefont {von Oppen}, \citenamefont
  {Glazman},\ and\ \citenamefont {Harris}}]{BJ2009}%
  \BibitemOpen
  \bibfield  {author} {\bibinfo {author} {\bibfnamefont {A.~C.}\ \bibnamefont
  {Bleszynski-Jayich}}, \bibinfo {author} {\bibfnamefont {W.~E.}\ \bibnamefont
  {Shanks}}, \bibinfo {author} {\bibfnamefont {B.}~\bibnamefont {Peaudecerf}},
  \bibinfo {author} {\bibfnamefont {E.}~\bibnamefont {Ginossar}}, \bibinfo
  {author} {\bibfnamefont {F.}~\bibnamefont {von Oppen}}, \bibinfo {author}
  {\bibfnamefont {L.}~\bibnamefont {Glazman}}, \ and\ \bibinfo {author}
  {\bibfnamefont {J.~G.~E.}\ \bibnamefont {Harris}},\ }\href {\doibase
  10.1126/science.1178139} {\bibfield  {journal} {\bibinfo  {journal}
  {Science}\ }\textbf {\bibinfo {volume} {326}},\ \bibinfo {pages} {272}
  (\bibinfo {year} {2009})}\BibitemShut {NoStop}%
\bibitem [{\citenamefont {Bluhm}\ \emph {et~al.}(2009)\citenamefont {Bluhm},
  \citenamefont {Koshnick}, \citenamefont {Bert}, \citenamefont {Huber},\ and\
  \citenamefont {Moler}}]{Bluhm2009}%
  \BibitemOpen
  \bibfield  {author} {\bibinfo {author} {\bibfnamefont {H.}~\bibnamefont
  {Bluhm}}, \bibinfo {author} {\bibfnamefont {N.~C.}\ \bibnamefont {Koshnick}},
  \bibinfo {author} {\bibfnamefont {J.~A.}\ \bibnamefont {Bert}}, \bibinfo
  {author} {\bibfnamefont {M.~E.}\ \bibnamefont {Huber}}, \ and\ \bibinfo
  {author} {\bibfnamefont {K.~A.}\ \bibnamefont {Moler}},\ }\href {\doibase
  10.1103/PhysRevLett.102.136802} {\bibfield  {journal} {\bibinfo  {journal}
  {Phys. Rev. Lett.}\ }\textbf {\bibinfo {volume} {102}},\ \bibinfo {pages}
  {136802} (\bibinfo {year} {2009})}\BibitemShut {NoStop}%
\bibitem [{\citenamefont {Souche}\ \emph {et~al.}(2013)\citenamefont {Souche},
  \citenamefont {Huillery}, \citenamefont {Pothier}, \citenamefont {Gandit},
  \citenamefont {Mars}, \citenamefont {Skipetrov},\ and\ \citenamefont
  {Bourgeois}}]{Souche2013}%
  \BibitemOpen
  \bibfield  {author} {\bibinfo {author} {\bibfnamefont {G.~M.}\ \bibnamefont
  {Souche}}, \bibinfo {author} {\bibfnamefont {J.}~\bibnamefont {Huillery}},
  \bibinfo {author} {\bibfnamefont {H.}~\bibnamefont {Pothier}}, \bibinfo
  {author} {\bibfnamefont {P.}~\bibnamefont {Gandit}}, \bibinfo {author}
  {\bibfnamefont {J.~I.}\ \bibnamefont {Mars}}, \bibinfo {author}
  {\bibfnamefont {S.~E.}\ \bibnamefont {Skipetrov}}, \ and\ \bibinfo {author}
  {\bibfnamefont {O.}~\bibnamefont {Bourgeois}},\ }\href {\doibase
  10.1103/PhysRevB.87.115120} {\bibfield  {journal} {\bibinfo  {journal} {Phys.
  Rev. B}\ }\textbf {\bibinfo {volume} {87}},\ \bibinfo {pages} {115120}
  (\bibinfo {year} {2013})}\BibitemShut {NoStop}%
\bibitem [{\citenamefont {Castro~Neto}\ \emph {et~al.}(2009)\citenamefont
  {Castro~Neto}, \citenamefont {Guinea}, \citenamefont {Peres}, \citenamefont
  {Novoselov},\ and\ \citenamefont {Geim}}]{CastroNeto2009}%
  \BibitemOpen
  \bibfield  {author} {\bibinfo {author} {\bibfnamefont {A.~H.}\ \bibnamefont
  {Castro~Neto}}, \bibinfo {author} {\bibfnamefont {F.}~\bibnamefont {Guinea}},
  \bibinfo {author} {\bibfnamefont {N.~M.~R.}\ \bibnamefont {Peres}}, \bibinfo
  {author} {\bibfnamefont {K.~S.}\ \bibnamefont {Novoselov}}, \ and\ \bibinfo
  {author} {\bibfnamefont {A.~K.}\ \bibnamefont {Geim}},\ }\href {\doibase
  10.1103/RevModPhys.81.109} {\bibfield  {journal} {\bibinfo  {journal} {Rev.
  Mod. Phys.}\ }\textbf {\bibinfo {volume} {81}},\ \bibinfo {pages} {109}
  (\bibinfo {year} {2009})}\BibitemShut {NoStop}%
\bibitem [{\citenamefont {Goerbig}(2011)}]{Goerbig2011}%
  \BibitemOpen
  \bibfield  {author} {\bibinfo {author} {\bibfnamefont {M.~O.}\ \bibnamefont
  {Goerbig}},\ }\href {\doibase 10.1103/RevModPhys.83.1193} {\bibfield
  {journal} {\bibinfo  {journal} {Rev. Mod. Phys.}\ }\textbf {\bibinfo {volume}
  {83}},\ \bibinfo {pages} {1193} (\bibinfo {year} {2011})}\BibitemShut
  {NoStop}%
\bibitem [{\citenamefont {Hasan}\ and\ \citenamefont {Kane}(2010)}]{Hasan2010}%
  \BibitemOpen
  \bibfield  {author} {\bibinfo {author} {\bibfnamefont {M.~Z.}\ \bibnamefont
  {Hasan}}\ and\ \bibinfo {author} {\bibfnamefont {C.~L.}\ \bibnamefont
  {Kane}},\ }\href {\doibase 10.1103/RevModPhys.82.3045} {\bibfield  {journal}
  {\bibinfo  {journal} {Rev. Mod. Phys.}\ }\textbf {\bibinfo {volume} {82}},\
  \bibinfo {pages} {3045} (\bibinfo {year} {2010})}\BibitemShut {NoStop}%
\bibitem [{\citenamefont {Qi}\ and\ \citenamefont {Zhang}(2011)}]{Qi2011}%
  \BibitemOpen
  \bibfield  {author} {\bibinfo {author} {\bibfnamefont {X.-L.}\ \bibnamefont
  {Qi}}\ and\ \bibinfo {author} {\bibfnamefont {S.-C.}\ \bibnamefont {Zhang}},\
  }\href {\doibase 10.1103/RevModPhys.83.1057} {\bibfield  {journal} {\bibinfo
  {journal} {Rev. Mod. Phys.}\ }\textbf {\bibinfo {volume} {83}},\ \bibinfo
  {pages} {1057} (\bibinfo {year} {2011})}\BibitemShut {NoStop}%
\bibitem [{\citenamefont {Recher}\ \emph {et~al.}(2007)\citenamefont {Recher},
  \citenamefont {Trauzettel}, \citenamefont {Rycerz}, \citenamefont {Blanter},
  \citenamefont {Beenakker},\ and\ \citenamefont {Morpurgo}}]{Recher2007}%
  \BibitemOpen
  \bibfield  {author} {\bibinfo {author} {\bibfnamefont {P.}~\bibnamefont
  {Recher}}, \bibinfo {author} {\bibfnamefont {B.}~\bibnamefont {Trauzettel}},
  \bibinfo {author} {\bibfnamefont {A.}~\bibnamefont {Rycerz}}, \bibinfo
  {author} {\bibfnamefont {Y.~M.}\ \bibnamefont {Blanter}}, \bibinfo {author}
  {\bibfnamefont {C.~W.~J.}\ \bibnamefont {Beenakker}}, \ and\ \bibinfo
  {author} {\bibfnamefont {A.~F.}\ \bibnamefont {Morpurgo}},\ }\href {\doibase
  10.1103/PhysRevB.76.235404} {\bibfield  {journal} {\bibinfo  {journal} {Phys.
  Rev. B}\ }\textbf {\bibinfo {volume} {76}},\ \bibinfo {pages} {235404}
  (\bibinfo {year} {2007})}\BibitemShut {NoStop}%
\bibitem [{\citenamefont {Zarenia}\ \emph {et~al.}(2010)\citenamefont
  {Zarenia}, \citenamefont {Pereira}, \citenamefont {Chaves}, \citenamefont
  {Peeters},\ and\ \citenamefont {Farias}}]{Zarenia2010}%
  \BibitemOpen
  \bibfield  {author} {\bibinfo {author} {\bibfnamefont {M.}~\bibnamefont
  {Zarenia}}, \bibinfo {author} {\bibfnamefont {J.~M.}\ \bibnamefont
  {Pereira}}, \bibinfo {author} {\bibfnamefont {A.}~\bibnamefont {Chaves}},
  \bibinfo {author} {\bibfnamefont {F.~M.}\ \bibnamefont {Peeters}}, \ and\
  \bibinfo {author} {\bibfnamefont {G.~A.}\ \bibnamefont {Farias}},\ }\href
  {\doibase 10.1103/PhysRevB.81.045431} {\bibfield  {journal} {\bibinfo
  {journal} {Phys. Rev. B}\ }\textbf {\bibinfo {volume} {81}},\ \bibinfo
  {pages} {045431} (\bibinfo {year} {2010})}\BibitemShut {NoStop}%
\bibitem [{\citenamefont {Michetti}\ and\ \citenamefont
  {Recher}(2011)}]{Michetti2011}%
  \BibitemOpen
  \bibfield  {author} {\bibinfo {author} {\bibfnamefont {P.}~\bibnamefont
  {Michetti}}\ and\ \bibinfo {author} {\bibfnamefont {P.}~\bibnamefont
  {Recher}},\ }\href {\doibase 10.1103/PhysRevB.83.125420} {\bibfield
  {journal} {\bibinfo  {journal} {Phys. Rev. B}\ }\textbf {\bibinfo {volume}
  {83}},\ \bibinfo {pages} {125420} (\bibinfo {year} {2011})}\BibitemShut
  {NoStop}%
\bibitem [{\citenamefont {Manton}(1985)}]{Manton1985}%
  \BibitemOpen
  \bibfield  {author} {\bibinfo {author} {\bibfnamefont {N.}~\bibnamefont
  {Manton}},\ }\href {\doibase 10.1016/0003-4916(85)90199-X} {\bibfield
  {journal} {\bibinfo  {journal} {Ann. Phys.}\ }\textbf {\bibinfo {volume}
  {159}},\ \bibinfo {pages} {220 } (\bibinfo {year} {1985})}\BibitemShut
  {NoStop}%
\bibitem [{\citenamefont {Shifman}(1991)}]{Shifman1991}%
  \BibitemOpen
  \bibfield  {author} {\bibinfo {author} {\bibfnamefont {M.}~\bibnamefont
  {Shifman}},\ }\href {\doibase 10.1016/0370-1573(91)90020-M} {\bibfield
  {journal} {\bibinfo  {journal} {Phys. Rep.}\ }\textbf {\bibinfo {volume}
  {209}},\ \bibinfo {pages} {341 } (\bibinfo {year} {1991})}\BibitemShut
  {NoStop}%
\bibitem [{\citenamefont {Kohno}\ \emph {et~al.}(1992)\citenamefont {Kohno},
  \citenamefont {Yoshioka},\ and\ \citenamefont {Fukuyama}}]{Kohno1992}%
  \BibitemOpen
  \bibfield  {author} {\bibinfo {author} {\bibfnamefont {H.}~\bibnamefont
  {Kohno}}, \bibinfo {author} {\bibfnamefont {H.}~\bibnamefont {Yoshioka}}, \
  and\ \bibinfo {author} {\bibfnamefont {H.}~\bibnamefont {Fukuyama}},\ }\href
  {\doibase 10.1143/JPSJ.61.3462} {\bibfield  {journal} {\bibinfo  {journal}
  {J. Phys. Soc. Jpn.}\ }\textbf {\bibinfo {volume} {61}},\ \bibinfo {pages}
  {3462} (\bibinfo {year} {1992})}\BibitemShut {NoStop}%
\bibitem [{\citenamefont {Loss}(1992)}]{Loss1992a}%
  \BibitemOpen
  \bibfield  {author} {\bibinfo {author} {\bibfnamefont {D.}~\bibnamefont
  {Loss}},\ }\href {\doibase 10.1103/PhysRevLett.69.343} {\bibfield  {journal}
  {\bibinfo  {journal} {Phys. Rev. Lett.}\ }\textbf {\bibinfo {volume} {69}},\
  \bibinfo {pages} {343} (\bibinfo {year} {1992})}\BibitemShut {NoStop}%
\bibitem [{\citenamefont {Gogolin}\ and\ \citenamefont
  {Prokof'ev}(1994)}]{Gogolin1994}%
  \BibitemOpen
  \bibfield  {author} {\bibinfo {author} {\bibfnamefont {A.~O.}\ \bibnamefont
  {Gogolin}}\ and\ \bibinfo {author} {\bibfnamefont {N.~V.}\ \bibnamefont
  {Prokof'ev}},\ }\href {\doibase 10.1103/PhysRevB.50.4921} {\bibfield
  {journal} {\bibinfo  {journal} {Phys. Rev. B}\ }\textbf {\bibinfo {volume}
  {50}},\ \bibinfo {pages} {4921} (\bibinfo {year} {1994})}\BibitemShut
  {NoStop}%
\bibitem [{\citenamefont {Affleck}\ and\ \citenamefont
  {Simon}(2001)}]{Affleck2001}%
  \BibitemOpen
  \bibfield  {author} {\bibinfo {author} {\bibfnamefont {I.}~\bibnamefont
  {Affleck}}\ and\ \bibinfo {author} {\bibfnamefont {P.}~\bibnamefont
  {Simon}},\ }\href {\doibase 10.1103/PhysRevLett.86.2854} {\bibfield
  {journal} {\bibinfo  {journal} {Phys. Rev. Lett.}\ }\textbf {\bibinfo
  {volume} {86}},\ \bibinfo {pages} {2854} (\bibinfo {year}
  {2001})}\BibitemShut {NoStop}%
\bibitem [{\citenamefont {Eckle}\ \emph {et~al.}(2001)\citenamefont {Eckle},
  \citenamefont {Johannesson},\ and\ \citenamefont {Stafford}}]{Eckle2001}%
  \BibitemOpen
  \bibfield  {author} {\bibinfo {author} {\bibfnamefont {H.-P.}\ \bibnamefont
  {Eckle}}, \bibinfo {author} {\bibfnamefont {H.}~\bibnamefont {Johannesson}},
  \ and\ \bibinfo {author} {\bibfnamefont {C.~A.}\ \bibnamefont {Stafford}},\
  }\href {\doibase 10.1103/PhysRevLett.87.016602} {\bibfield  {journal}
  {\bibinfo  {journal} {Phys. Rev. Lett.}\ }\textbf {\bibinfo {volume} {87}},\
  \bibinfo {pages} {016602} (\bibinfo {year} {2001})}\BibitemShut {NoStop}%
\bibitem [{\citenamefont {Peng}\ \emph {et~al.}(2010)\citenamefont {Peng},
  \citenamefont {Lai}, \citenamefont {Kong}, \citenamefont {Meister},
  \citenamefont {Chen}, \citenamefont {Qi}, \citenamefont {Zhang},
  \citenamefont {Shen},\ and\ \citenamefont {Cui}}]{Peng2010}%
  \BibitemOpen
  \bibfield  {author} {\bibinfo {author} {\bibfnamefont {H.}~\bibnamefont
  {Peng}}, \bibinfo {author} {\bibfnamefont {K.}~\bibnamefont {Lai}}, \bibinfo
  {author} {\bibfnamefont {D.}~\bibnamefont {Kong}}, \bibinfo {author}
  {\bibfnamefont {S.}~\bibnamefont {Meister}}, \bibinfo {author} {\bibfnamefont
  {Y.}~\bibnamefont {Chen}}, \bibinfo {author} {\bibfnamefont {X.-L.}\
  \bibnamefont {Qi}}, \bibinfo {author} {\bibfnamefont {S.-C.}\ \bibnamefont
  {Zhang}}, \bibinfo {author} {\bibfnamefont {Z.-X.}\ \bibnamefont {Shen}}, \
  and\ \bibinfo {author} {\bibfnamefont {Y.}~\bibnamefont {Cui}},\ }\href
  {http://dx.doi.org/10.1038/nmat2609} {\bibfield  {journal} {\bibinfo
  {journal} {Nat. Mater.}\ }\textbf {\bibinfo {volume} {9}},\ \bibinfo {pages}
  {225} (\bibinfo {year} {2010})}\BibitemShut {NoStop}%
\bibitem [{\citenamefont {Dufouleur}\ \emph {et~al.}(2013)\citenamefont
  {Dufouleur}, \citenamefont {Veyrat}, \citenamefont {Teichgr\"aber},
  \citenamefont {Neuhaus}, \citenamefont {Nowka}, \citenamefont {Hampel},
  \citenamefont {Cayssol}, \citenamefont {Schumann}, \citenamefont {Eichler},
  \citenamefont {Schmidt}, \citenamefont {B\"uchner},\ and\ \citenamefont
  {Giraud}}]{Dufouleur2013}%
  \BibitemOpen
  \bibfield  {author} {\bibinfo {author} {\bibfnamefont {J.}~\bibnamefont
  {Dufouleur}}, \bibinfo {author} {\bibfnamefont {L.}~\bibnamefont {Veyrat}},
  \bibinfo {author} {\bibfnamefont {A.}~\bibnamefont {Teichgr\"aber}}, \bibinfo
  {author} {\bibfnamefont {S.}~\bibnamefont {Neuhaus}}, \bibinfo {author}
  {\bibfnamefont {C.}~\bibnamefont {Nowka}}, \bibinfo {author} {\bibfnamefont
  {S.}~\bibnamefont {Hampel}}, \bibinfo {author} {\bibfnamefont
  {J.}~\bibnamefont {Cayssol}}, \bibinfo {author} {\bibfnamefont
  {J.}~\bibnamefont {Schumann}}, \bibinfo {author} {\bibfnamefont
  {B.}~\bibnamefont {Eichler}}, \bibinfo {author} {\bibfnamefont {O.~G.}\
  \bibnamefont {Schmidt}}, \bibinfo {author} {\bibfnamefont {B.}~\bibnamefont
  {B\"uchner}}, \ and\ \bibinfo {author} {\bibfnamefont {R.}~\bibnamefont
  {Giraud}},\ }\href {\doibase 10.1103/PhysRevLett.110.186806} {\bibfield
  {journal} {\bibinfo  {journal} {Phys. Rev. Lett.}\ }\textbf {\bibinfo
  {volume} {110}},\ \bibinfo {pages} {186806} (\bibinfo {year}
  {2013})}\BibitemShut {NoStop}%
\bibitem [{\citenamefont {Bardarson}\ \emph {et~al.}(2010)\citenamefont
  {Bardarson}, \citenamefont {Brouwer},\ and\ \citenamefont
  {Moore}}]{Bardarson2010}%
  \BibitemOpen
  \bibfield  {author} {\bibinfo {author} {\bibfnamefont {J.~H.}\ \bibnamefont
  {Bardarson}}, \bibinfo {author} {\bibfnamefont {P.~W.}\ \bibnamefont
  {Brouwer}}, \ and\ \bibinfo {author} {\bibfnamefont {J.~E.}\ \bibnamefont
  {Moore}},\ }\href {\doibase 10.1103/PhysRevLett.105.156803} {\bibfield
  {journal} {\bibinfo  {journal} {Phys. Rev. Lett.}\ }\textbf {\bibinfo
  {volume} {105}},\ \bibinfo {pages} {156803} (\bibinfo {year}
  {2010})}\BibitemShut {NoStop}%
\bibitem [{\citenamefont {Zhang}\ and\ \citenamefont
  {Vishwanath}(2010)}]{Zhang2010}%
  \BibitemOpen
  \bibfield  {author} {\bibinfo {author} {\bibfnamefont {Y.}~\bibnamefont
  {Zhang}}\ and\ \bibinfo {author} {\bibfnamefont {A.}~\bibnamefont
  {Vishwanath}},\ }\href {\doibase 10.1103/PhysRevLett.105.206601} {\bibfield
  {journal} {\bibinfo  {journal} {Phys. Rev. Lett.}\ }\textbf {\bibinfo
  {volume} {105}},\ \bibinfo {pages} {206601} (\bibinfo {year}
  {2010})}\BibitemShut {NoStop}%
\bibitem [{\citenamefont {Bardarson}\ and\ \citenamefont
  {Moore}(2013)}]{Bardarson2013}%
  \BibitemOpen
  \bibfield  {author} {\bibinfo {author} {\bibfnamefont {J.~H.}\ \bibnamefont
  {Bardarson}}\ and\ \bibinfo {author} {\bibfnamefont {J.~E.}\ \bibnamefont
  {Moore}},\ }\href {http://stacks.iop.org/0034-4885/76/i=5/a=056501}
  {\bibfield  {journal} {\bibinfo  {journal} {Rep. Prog. Phys.}\ }\textbf
  {\bibinfo {volume} {76}},\ \bibinfo {pages} {056501} (\bibinfo {year}
  {2013})}\BibitemShut {NoStop}%
\bibitem [{\citenamefont {Imura}\ \emph {et~al.}(2012)\citenamefont {Imura},
  \citenamefont {Yoshimura}, \citenamefont {Takane},\ and\ \citenamefont
  {Fukui}}]{Imura2012}%
  \BibitemOpen
  \bibfield  {author} {\bibinfo {author} {\bibfnamefont {K.-I.}\ \bibnamefont
  {Imura}}, \bibinfo {author} {\bibfnamefont {Y.}~\bibnamefont {Yoshimura}},
  \bibinfo {author} {\bibfnamefont {Y.}~\bibnamefont {Takane}}, \ and\ \bibinfo
  {author} {\bibfnamefont {T.}~\bibnamefont {Fukui}},\ }\href {\doibase
  10.1103/PhysRevB.86.235119} {\bibfield  {journal} {\bibinfo  {journal} {Phys.
  Rev. B}\ }\textbf {\bibinfo {volume} {86}},\ \bibinfo {pages} {235119}
  (\bibinfo {year} {2012})}\BibitemShut {NoStop}%
\bibitem [{\citenamefont {Creutz}\ and\ \citenamefont
  {Horv\'ath}(1994)}]{Creutz1994}%
  \BibitemOpen
  \bibfield  {author} {\bibinfo {author} {\bibfnamefont {M.}~\bibnamefont
  {Creutz}}\ and\ \bibinfo {author} {\bibfnamefont {I.}~\bibnamefont
  {Horv\'ath}},\ }\href {\doibase 10.1103/PhysRevD.50.2297} {\bibfield
  {journal} {\bibinfo  {journal} {Phys. Rev. D}\ }\textbf {\bibinfo {volume}
  {50}},\ \bibinfo {pages} {2297} (\bibinfo {year} {1994})}\BibitemShut
  {NoStop}%
\bibitem [{\citenamefont {Creutz}(1999)}]{Creutz1999}%
  \BibitemOpen
  \bibfield  {author} {\bibinfo {author} {\bibfnamefont {M.}~\bibnamefont
  {Creutz}},\ }\href {\doibase 10.1103/PhysRevLett.83.2636} {\bibfield
  {journal} {\bibinfo  {journal} {Phys. Rev. Lett.}\ }\textbf {\bibinfo
  {volume} {83}},\ \bibinfo {pages} {2636} (\bibinfo {year}
  {1999})}\BibitemShut {NoStop}%
\bibitem [{\citenamefont {Bermudez}\ \emph {et~al.}(2009)\citenamefont
  {Bermudez}, \citenamefont {Patan\`e}, \citenamefont {Amico},\ and\
  \citenamefont {Martin-Delgado}}]{Bermudez2009}%
  \BibitemOpen
  \bibfield  {author} {\bibinfo {author} {\bibfnamefont {A.}~\bibnamefont
  {Bermudez}}, \bibinfo {author} {\bibfnamefont {D.}~\bibnamefont {Patan\`e}},
  \bibinfo {author} {\bibfnamefont {L.}~\bibnamefont {Amico}}, \ and\ \bibinfo
  {author} {\bibfnamefont {M.~A.}\ \bibnamefont {Martin-Delgado}},\ }\href
  {\doibase 10.1103/PhysRevLett.102.135702} {\bibfield  {journal} {\bibinfo
  {journal} {Phys. Rev. Lett.}\ }\textbf {\bibinfo {volume} {102}},\ \bibinfo
  {pages} {135702} (\bibinfo {year} {2009})}\BibitemShut {NoStop}%
\bibitem [{\citenamefont {Viyuela}\ \emph {et~al.}(2012)\citenamefont
  {Viyuela}, \citenamefont {Rivas},\ and\ \citenamefont
  {Martin-Delgado}}]{Viyuela2012}%
  \BibitemOpen
  \bibfield  {author} {\bibinfo {author} {\bibfnamefont {O.}~\bibnamefont
  {Viyuela}}, \bibinfo {author} {\bibfnamefont {A.}~\bibnamefont {Rivas}}, \
  and\ \bibinfo {author} {\bibfnamefont {M.~A.}\ \bibnamefont
  {Martin-Delgado}},\ }\href {\doibase 10.1103/PhysRevB.86.155140} {\bibfield
  {journal} {\bibinfo  {journal} {Phys. Rev. B}\ }\textbf {\bibinfo {volume}
  {86}},\ \bibinfo {pages} {155140} (\bibinfo {year} {2012})}\BibitemShut
  {NoStop}%
\bibitem [{\citenamefont {Mazza}\ \emph {et~al.}(2012)\citenamefont {Mazza},
  \citenamefont {Bermudez}, \citenamefont {Goldman}, \citenamefont {Rizzi},
  \citenamefont {Martin-Delgado},\ and\ \citenamefont
  {Lewenstein}}]{Mazza2012}%
  \BibitemOpen
  \bibfield  {author} {\bibinfo {author} {\bibfnamefont {L.}~\bibnamefont
  {Mazza}}, \bibinfo {author} {\bibfnamefont {A.}~\bibnamefont {Bermudez}},
  \bibinfo {author} {\bibfnamefont {N.}~\bibnamefont {Goldman}}, \bibinfo
  {author} {\bibfnamefont {M.}~\bibnamefont {Rizzi}}, \bibinfo {author}
  {\bibfnamefont {M.~A.}\ \bibnamefont {Martin-Delgado}}, \ and\ \bibinfo
  {author} {\bibfnamefont {M.}~\bibnamefont {Lewenstein}},\ }\href@noop {}
  {\bibfield  {journal} {\bibinfo  {journal} {New Journal of Physics}\ }\textbf
  {\bibinfo {volume} {14}},\ \bibinfo {pages} {015007} (\bibinfo {year}
  {2012})}\BibitemShut {NoStop}%
\bibitem [{\citenamefont {Schnyder}\ \emph {et~al.}(2008)\citenamefont
  {Schnyder}, \citenamefont {Ryu}, \citenamefont {Furusaki},\ and\
  \citenamefont {Ludwig}}]{Schnyder2008}%
  \BibitemOpen
  \bibfield  {author} {\bibinfo {author} {\bibfnamefont {A.~P.}\ \bibnamefont
  {Schnyder}}, \bibinfo {author} {\bibfnamefont {S.}~\bibnamefont {Ryu}},
  \bibinfo {author} {\bibfnamefont {A.}~\bibnamefont {Furusaki}}, \ and\
  \bibinfo {author} {\bibfnamefont {A.~W.~W.}\ \bibnamefont {Ludwig}},\ }\href
  {\doibase 10.1103/PhysRevB.78.195125} {\bibfield  {journal} {\bibinfo
  {journal} {Phys. Rev. B}\ }\textbf {\bibinfo {volume} {78}},\ \bibinfo
  {pages} {195125} (\bibinfo {year} {2008})}\BibitemShut {NoStop}%
\bibitem [{\citenamefont {Kitaev}(2009)}]{Kitaev2009}%
  \BibitemOpen
  \bibfield  {author} {\bibinfo {author} {\bibfnamefont {A.}~\bibnamefont
  {Kitaev}},\ }\href {\doibase 10.1063/1.3149495} {\bibfield  {journal}
  {\bibinfo  {journal} {AIP Conf. Proc.}\ }\textbf {\bibinfo {volume} {1134}},\
  \bibinfo {pages} {22} (\bibinfo {year} {2009})}\BibitemShut {NoStop}%
\bibitem [{\citenamefont {Ryu}\ \emph {et~al.}(2010)\citenamefont {Ryu},
  \citenamefont {Schnyder}, \citenamefont {Furusaki},\ and\ \citenamefont
  {Ludwig}}]{Ryu2010}%
  \BibitemOpen
  \bibfield  {author} {\bibinfo {author} {\bibfnamefont {S.}~\bibnamefont
  {Ryu}}, \bibinfo {author} {\bibfnamefont {A.~P.}\ \bibnamefont {Schnyder}},
  \bibinfo {author} {\bibfnamefont {A.}~\bibnamefont {Furusaki}}, \ and\
  \bibinfo {author} {\bibfnamefont {A.~W.~W.}\ \bibnamefont {Ludwig}},\ }\href
  {http://stacks.iop.org/1367-2630/12/i=6/a=065010} {\bibfield  {journal}
  {\bibinfo  {journal} {New J. Phys.}\ }\textbf {\bibinfo {volume} {12}},\
  \bibinfo {pages} {065010} (\bibinfo {year} {2010})}\BibitemShut {NoStop}%
\bibitem [{\citenamefont {Dirac}(1928)}]{Dirac1928}%
  \BibitemOpen
  \bibfield  {author} {\bibinfo {author} {\bibfnamefont {P.~A.~M.}\
  \bibnamefont {Dirac}},\ }\href {http://www.jstor.org/stable/94981} {\bibfield
   {journal} {\bibinfo  {journal} {P. R. Soc. Lond. A-Conta.}\ }\textbf
  {\bibinfo {volume} {117}},\ \bibinfo {pages} {pp. 610} (\bibinfo {year}
  {1928})}\BibitemShut {NoStop}%
\bibitem [{\citenamefont {Dirac}(1930)}]{Dirac1930}%
  \BibitemOpen
  \bibfield  {author} {\bibinfo {author} {\bibfnamefont {P.~A.~M.}\
  \bibnamefont {Dirac}},\ }\href {\doibase 10.1098/rspa.1930.0013} {\bibfield
  {journal} {\bibinfo  {journal} {Proc. R. Soc. Lon. Ser.-A}\ }\textbf
  {\bibinfo {volume} {126}},\ \bibinfo {pages} {360} (\bibinfo {year}
  {1930})}\BibitemShut {NoStop}%
\bibitem [{\citenamefont {Bardeen}\ and\ \citenamefont
  {Johnson}(1972)}]{Bardeen:1972}%
  \BibitemOpen
  \bibfield  {author} {\bibinfo {author} {\bibfnamefont {J.}~\bibnamefont
  {Bardeen}}\ and\ \bibinfo {author} {\bibfnamefont {J.~L.}\ \bibnamefont
  {Johnson}},\ }\href {\doibase 10.1103/PhysRevB.5.72} {\bibfield  {journal}
  {\bibinfo  {journal} {Phys. Rev. B}\ }\textbf {\bibinfo {volume} {5}},\
  \bibinfo {pages} {72} (\bibinfo {year} {1972})}\BibitemShut {NoStop}%
\bibitem [{\citenamefont {Affleck}\ \emph {et~al.}(2000)\citenamefont
  {Affleck}, \citenamefont {Caux},\ and\ \citenamefont
  {Zagoskin}}]{Affleck:2000}%
  \BibitemOpen
  \bibfield  {author} {\bibinfo {author} {\bibfnamefont {I.}~\bibnamefont
  {Affleck}}, \bibinfo {author} {\bibfnamefont {J.-S.}\ \bibnamefont {Caux}}, \
  and\ \bibinfo {author} {\bibfnamefont {A.~M.}\ \bibnamefont {Zagoskin}},\
  }\href {\doibase 10.1103/PhysRevB.62.1433} {\bibfield  {journal} {\bibinfo
  {journal} {Phys. Rev. B}\ }\textbf {\bibinfo {volume} {62}},\ \bibinfo
  {pages} {1433} (\bibinfo {year} {2000})}\BibitemShut {NoStop}%
\bibitem [{\citenamefont {Cayssol}\ \emph {et~al.}(2003)\citenamefont
  {Cayssol}, \citenamefont {Kontos},\ and\ \citenamefont
  {Montambaux}}]{Cayssol:2003}%
  \BibitemOpen
  \bibfield  {author} {\bibinfo {author} {\bibfnamefont {J.}~\bibnamefont
  {Cayssol}}, \bibinfo {author} {\bibfnamefont {T.}~\bibnamefont {Kontos}}, \
  and\ \bibinfo {author} {\bibfnamefont {G.}~\bibnamefont {Montambaux}},\
  }\href {\doibase 10.1103/PhysRevB.67.184508} {\bibfield  {journal} {\bibinfo
  {journal} {Phys. Rev. B}\ }\textbf {\bibinfo {volume} {67}},\ \bibinfo
  {pages} {184508} (\bibinfo {year} {2003})}\BibitemShut {NoStop}%
\bibitem [{\citenamefont {Gogolin}\ \emph {et~al.}(2004)\citenamefont
  {Gogolin}, \citenamefont {Nersesyan},\ and\ \citenamefont
  {Tsvelik}}]{Gogolin2004}%
  \BibitemOpen
  \bibfield  {author} {\bibinfo {author} {\bibfnamefont {A.~O.}\ \bibnamefont
  {Gogolin}}, \bibinfo {author} {\bibfnamefont {A.~A.}\ \bibnamefont
  {Nersesyan}}, \ and\ \bibinfo {author} {\bibfnamefont {A.~M.}\ \bibnamefont
  {Tsvelik}},\ }\href {http://amazon.com/o/ASIN/0521617197/} {\emph {\bibinfo
  {title} {Bosonization and Strongly Correlated Systems}}}\ (\bibinfo
  {publisher} {Cambridge University Press},\ \bibinfo {year}
  {2004})\BibitemShut {NoStop}%
\bibitem [{\citenamefont {Kravtsov}\ and\ \citenamefont
  {Zirnbauer}(1992)}]{Kravtsov1992}%
  \BibitemOpen
  \bibfield  {author} {\bibinfo {author} {\bibfnamefont {V.~E.}\ \bibnamefont
  {Kravtsov}}\ and\ \bibinfo {author} {\bibfnamefont {M.~R.}\ \bibnamefont
  {Zirnbauer}},\ }\href {\doibase 10.1103/PhysRevB.46.4332} {\bibfield
  {journal} {\bibinfo  {journal} {Phys. Rev. B}\ }\textbf {\bibinfo {volume}
  {46}},\ \bibinfo {pages} {4332} (\bibinfo {year} {1992})}\BibitemShut
  {NoStop}%
\bibitem [{\citenamefont {Ilan}\ \emph {et~al.}(2012)\citenamefont {Ilan},
  \citenamefont {Cayssol}, \citenamefont {Bardarson},\ and\ \citenamefont
  {Moore}}]{Ilan2012}%
  \BibitemOpen
  \bibfield  {author} {\bibinfo {author} {\bibfnamefont {R.}~\bibnamefont
  {Ilan}}, \bibinfo {author} {\bibfnamefont {J.}~\bibnamefont {Cayssol}},
  \bibinfo {author} {\bibfnamefont {J.~H.}\ \bibnamefont {Bardarson}}, \ and\
  \bibinfo {author} {\bibfnamefont {J.~E.}\ \bibnamefont {Moore}},\ }\href
  {\doibase 10.1103/PhysRevLett.109.216602} {\bibfield  {journal} {\bibinfo
  {journal} {Phys. Rev. Lett.}\ }\textbf {\bibinfo {volume} {109}},\ \bibinfo
  {pages} {216602} (\bibinfo {year} {2012})}\BibitemShut {NoStop}%
\bibitem [{\citenamefont {B\"uttiker}(1985)}]{Buettiker1985}%
  \BibitemOpen
  \bibfield  {author} {\bibinfo {author} {\bibfnamefont {M.}~\bibnamefont
  {B\"uttiker}},\ }\href {\doibase 10.1103/PhysRevB.32.1846} {\bibfield
  {journal} {\bibinfo  {journal} {Phys. Rev. B}\ }\textbf {\bibinfo {volume}
  {32}},\ \bibinfo {pages} {1846} (\bibinfo {year} {1985})}\BibitemShut
  {NoStop}%
\bibitem [{\citenamefont {Moskalets}(2010)}]{Moskalets2010}%
  \BibitemOpen
  \bibfield  {author} {\bibinfo {author} {\bibfnamefont {M.}~\bibnamefont
  {Moskalets}},\ }\href {\doibase 10.1063/1.3521568} {\bibfield  {journal}
  {\bibinfo  {journal} {Low Temp. Phys.}\ }\textbf {\bibinfo {volume} {36}},\
  \bibinfo {pages} {982} (\bibinfo {year} {2010})}\BibitemShut {NoStop}%
\bibitem [{\citenamefont {Wimmer}\ \emph {et~al.}(2010)\citenamefont {Wimmer},
  \citenamefont {Akhmerov}, \citenamefont {Medvedyeva}, \citenamefont
  {Tworzyd\l{}o},\ and\ \citenamefont {Beenakker}}]{Wimmer2010}%
  \BibitemOpen
  \bibfield  {author} {\bibinfo {author} {\bibfnamefont {M.}~\bibnamefont
  {Wimmer}}, \bibinfo {author} {\bibfnamefont {A.~R.}\ \bibnamefont
  {Akhmerov}}, \bibinfo {author} {\bibfnamefont {M.~V.}\ \bibnamefont
  {Medvedyeva}}, \bibinfo {author} {\bibfnamefont {J.}~\bibnamefont
  {Tworzyd\l{}o}}, \ and\ \bibinfo {author} {\bibfnamefont {C.~W.~J.}\
  \bibnamefont {Beenakker}},\ }\href {\doibase 10.1103/PhysRevLett.105.046803}
  {\bibfield  {journal} {\bibinfo  {journal} {Phys. Rev. Lett.}\ }\textbf
  {\bibinfo {volume} {105}},\ \bibinfo {pages} {046803} (\bibinfo {year}
  {2010})}\BibitemShut {NoStop}%
\end{thebibliography}%
\end{document}